\documentclass[review,3p]{elsarticle}

\usepackage{graphicx}
\usepackage[table]{xcolor}%
\usepackage{amssymb}
\usepackage{caption}
\usepackage{subcaption}
\usepackage{rotating}
\usepackage{float}
\usepackage{acronym}
\usepackage{booktabs}
\usepackage{titlesec}
\usepackage{color,soul}
\titleformat{\paragraph}
{\normalfont\normalsize\itshape}{\theparagraph}{1em}{}
\titlespacing*{\paragraph}
{0pt}{3.25ex plus 1ex minus .2ex}{1.5ex plus .2ex}

\usepackage{amsmath}
\usepackage[super]{nth}
\usepackage{longtable}
\usepackage{tocloft}
\usepackage{multirow}
\usepackage{textalpha}

\usepackage{array}
\newcolumntype{L}{>{\centering\arraybackslash}m{3cm}}

\usepackage[colorlinks=true]{hyperref}
\biboptions{authoryear}

\acrodef{AUC}{Area Under the Curve }
\acrodef{AWS}{Amazon Web Services}

\acrodef{BDAP}{“Big Data Analytics Platform”}
\acrodef{BRP}{'Basisregistratie Gewaspercelen'}
\acrodef{BSO}{Bare Soil}

\acrodef{CNN}{Convolutional Neural Networks}

\acrodef{F1}{F1 Score}
\acrodef{FPR}{False Positive Rate}

\acrodef{GEO}{Group on Earth Observation}

\acrodef{JEODPP}{Joint Research Center Earth Observation Data and Processing Platform}
\acrodef{JECAM}{Joint Experiment of Crop Assessment and Monitoring}
\acrodef{JRC}{Joint Research Centre}

\acrodef{M-F1}{Macro F1 Score}

\acrodef{PA}{Producer Accuracy}

\acrodef{UA}{User Accuracy}

\acrodef{ROC}{Receiver-Operating Curve}

\acrodef{SLI}{Street Level Imagery}

\acrodef{TPR}{True Positive Rate}

\acrodef{PPP}{Pictures Per Parcel}

\acrodef{CAR}{Carrot}
\acrodef{GMA}{Green Manure}
\acrodef{GRA}{Grassland}
\acrodef{GRS}{Grass Seeds}
\acrodef{MAI}{Maize}
\acrodef{ONI}{Onion}
\acrodef{POT}{Potato}
\acrodef{SBA}{Summer Barley}
\acrodef{SBT}{Sugar Beet}
\acrodef{SCR}{Spring Cereals}
\acrodef{SWH}{Spring Wheat}
\acrodef{TLP}{Tulip}
\acrodef{VEG}{Vegetable}
\acrodef{WBA}{Winter Barley}
\acrodef{WCR}{Winter Cereals}
\acrodef{WWH}{Winter Wheat}
\acrodef{OTH}{Other}

\begin{document}


\begin{frontmatter}
\title{Earth Observation based multi-scale analysis of crop diversity in the European Union: first insights for agro-environmental policies}



\author{Melissande Machefer$^{1,*}$, Matteo Zampieri\texorpdfstring{$^{1 2,*}$} ,  Marijn van der Velde \texorpdfstring{$^{1}$}, Frank Dentener \texorpdfstring{$^{1}$}, Martin Claverie \texorpdfstring{$^{1}$}, Rapha\"{e}l d'Andrimont \texorpdfstring{$^{1}$}}
\address{ $^{1}$\quad European Commission, Joint Research Centre (JRC), Ispra, Italy \\
$^{2}$\quad now at King Abdullah University of Science and Technology (KAUST), Thuwal, Saudi Arabia \\
 $^{*}$ Both authors have contributed equally.\\
Corresponding authors: melissande.machefer@ec.europa.eu, raphael.dandrimont@ec.europa.eu.
}


\begin{abstract}

Characterizing and quantifying crop diversity (“effective number  of crops”) across scales is needed to understand a wide range of issues related to resilience of farms and the agricultural sector, the provision of ecosystem services, and ultimately to provide a scientific basis for effective agro-environmental policies. We use a novel European Union (EU) wide satellite-derived product at 10 m spatial resolution to produce datasets of crop diversity across spatial (1-100 km) and administrative scales for the year 2018. We focus on the 27 EU countries and the United Kingdom. We define local crop diversity (α-diversity) at the 1km scale corresponding  to large farms or clusters of small-to-medium sized farms. 
Across countries, the α crop diversity  ranges from 2.3 to 4.4 with the highest levels achieved by systems dominated by a high number of small farms (less than 10 ha on average). Computed at grid level aggregation, γ-diversity (the number and area of crops that are grown independently from the precise location, for landscape region, and country levels) increases rapidly from 2.85 at 1 km to 3.86 at 10 km and levels off 4.27 at 100 km. Such diversity levels are higher than that reported for the U.S.A., likely related to differences in farm structure and practices. β-diversity,  the ratio of γ and α diversity, provides a measure of the diversity between agroecosystems and ranges from 1.2 to 2.3 across EU countries. Based on the magnitude and change of γ-diversity across scales, we classify countries’ diversity in four groups with possible consequences for regional to national agro-environmental policy recommendations, in particular the monitoring activities and indicator development of interventions for the implementation of the Common Agricultural Policy (CAP) in the EU. Forthcoming annual high-resolution continental Copernicus crop type maps will facilitate temporal comparisons. Various ecosystem co-variates are to be explored for deeper understanding of the link of crop diversity to agro-ecosystem services.


\end{abstract}


\end{frontmatter}

\renewcommand{\thetable}{Table \arabic{table}}
\renewcommand{\thefigure}{Fig. \arabic{figure}}
\renewcommand{\figurename}{}
\renewcommand{\tablename}{}

\section{Introduction}

Bolstering production resilience is essential in the face of climate change, and crop diversity is a key facet of production resilience \citep{elmqvist2003response,zampieri2020climate}. The stability of food production benefits from a diversity of crops grown \citep{elmqvist2003response,zampieri2020climate}, while crop diversity also promotes ecosystems services such as pest regulation \citep{thomine2022using}, soil biodiversity \citep{sprunger2020systems} and pollination \citep{raderschall2021landscape}. 

In agricultural policies that seek to support more sustainable practices, increasing crop diversity is often a key element. In the European Union (EU), the Common Agricultural Policy (CAP) introduced a voluntary crop \textit{diversification} scheme in 2013 in relation to the so-called greening payment (Regulation (EU) No 1306/2013). In the current CAP (2023-2027),  subsidies have become conditional on crop rotation rules following the introduction of a new environmental scheme called GAEC 7 (Good Agricultural and Environmental Conditions). 

 At the scale of farm holdings (0.01-100  ha, corresponding to spatial scales of  100 m - 1 km), crop diversification can trigger positive impacts on overall farm holding production and income \citep{davis2012increasing,gil2017resilience,hajjar2008utility,martin2015agricultural,prober2009enhancing,smith2008effects,wood2015functional}. In addition, positive relationships have been found between smaller farm sizes and crop diversity \citep{lazikova2019crop}. At national scale, several studies \citep{kahiluoto2019decline,renard2019national,zampieri2020estimating} find that crop diversity increases total production stability. This stability is especially pronounced if individual crops have diverse responses to climate anomalies, or if they are cultivated in varying agro-climatic regions within a country.

Climate change adaptation, but also conservation and restoration efforts, require an understanding of how crop diversity affects agro-ecosystem functioning on different scales ranging from field to farm up to continental scales. This requires robust metrics to quantify crop diversity \citep{aramburu2020scale,krishnaswamy2009quantifying,wood2015functional}.  
In the EU, crop diversity has been computed for several regions and countries using either farmers' declarations at parcel level, census data, or subnational crop area statistics, but understanding of their relationship needs to be improved. For instance, farmers' parcel level crop declarations included in the Land Parcel Identification System (LPIS) of the Integrated Administration and Control System (IACS) were used by \citet{uthes2020farm} to compute the Shannon index in the federal state of Brandenburg, Germany for the year 2017. Similarly, \citet{donfouet2017crop} and \citet{schaak2023long} use detailed LPIS data respectively for France in 2007 and Sweden for 2001-2018. \citet{bohan2021designing} used farmers declarations from three countries (Denmark, England and France) and computed crop richness for 12 crop categories in 5 km grid cells over several years of crop rotation. The Shannon Diversity Index was calculated by \citet{mahy2015simulating} for the Flanders region based on 24.839 farmers declarations in 2012. Only \citet{egli2021crop} present pan-EU results, using subnational crop-specific harvested areas derived from the EU's statistical office (EUROSTAT) to compute the Shannon Entropy Index. Using a survey of 79.532 farms in Bavaria conducted for the year 2010 for 41 crop categories, \citet{weigel2018crop} compute a modified Shannon index. While several indices have been developed to measure crop diversity, with respect to species' richness and evenness of species' proportional abundance, most of the above studies use the Shannon index. \citet{BONNEUIL2012280} discuss the scope and limits of the Shannon \citep{shannon1948mathematical}, Margalef \citep{margalef1973information}, and Simpson \citep{simpson1949measurement} crop diversity indices, where the Shannon Index is more sensitive to the presence of rare species. None of these studies with EU coverage have used high-resolution information from Earth Observation data. In contrast, in the USA, \citet{aramburu2020scale} have computed the Shannon Entropy Index from remotely sensed crop maps.

In this work, we enhance the assessment of crop diversity analysis in both spatial coverage and resolution by taking advantage of a recently published EU-wide 10-m resolution Earth Observation derived crop map \citep{d2021parcel} as described in section \ref{sec:method}. In section  \ref{sec:results}, our study comprehensively evaluates crop diversity across scales (from 1 to 100 km) with an analysis ranging from typical farm sizes within the EU, to European administrative delineations, offering insights in the context of agricultural policies. Furthermore, we explore the relationship between average national farm holding sizes and crop diversification levels and compare our results in the EU with the USA  \citep{aramburu2020scale}. In section \ref{sec:discussion}, we delve into the nuances, strengths, and challenges of employing such diversity indices as reliable measures of agricultural resilience, particularly within the framework of the CAP.

\section{Materials and methods} 
\label{sec:method}
\subsection{Study area}

Europe exhibits a remarkable diversity of climate types, owing to its vast geographical extent and complex topography \citep{beck2018present}. From the Mediterranean region in the south with its warm, dry and cloud-free summers and mild, wet winters to the subarctic climates of Scandinavia with long, harsh winters and short, cool summers, and everything in between, including temperate, continental, and oceanic climates. This climatic variation greatly influences agriculture across the continent. In the Mediterranean, crops like olives, grapes, and citrus fruits thrive in the warm, sunny summers. In contrast, cooler northern regions favor crops such as wheat, barley, and root vegetables. Additionally, the temperate climates of Western Europe support a wide range of crops including corn, rapeseed, and various fruits. Clearly, temperature and sunlight limitations in the northern countries strongly constrain the number of crops that can potentially be grown at the higher latitudes in Europe.

\subsection{Crop type map derived from Earth Observation} 

The 10-m resolution EU crop map was generated using satellite Sentinel-1 (S1) Synthetic Aperture Radar (SAR) observations from 2018, in combination with EUROSTAT LUCAS in-situ data as a training dataset. Algorithms trained with the S1 data time series capture the coverage of crops during the main growing season (January to end of July) on any agricultural area. As a result, every field in the EU cropped with wheat, maize, rapeseed, barley, potatoes, sugar beets, and other crop types (19 in total, see \ref{tab:eucropmap_class}) is mapped at a very fine 10 m x 10 m spatial scale, and used to derive EU-wide information on crop diversity.

The 80-85\% map accuracy reported by \citet{d2021parcel} is acceptable for the application of our study, since most classification errors are related to discriminating crops with similar growth characteristics such as common and durum wheat. In our analysis, we retain the information on all crop species allowing for follow-up analysis.

\subsection{Administrative units and grids} 
\label{subsec:admin_grids_units}

\begin{figure*}[!t]
    \centering 
    \includegraphics[width=.7\textwidth]{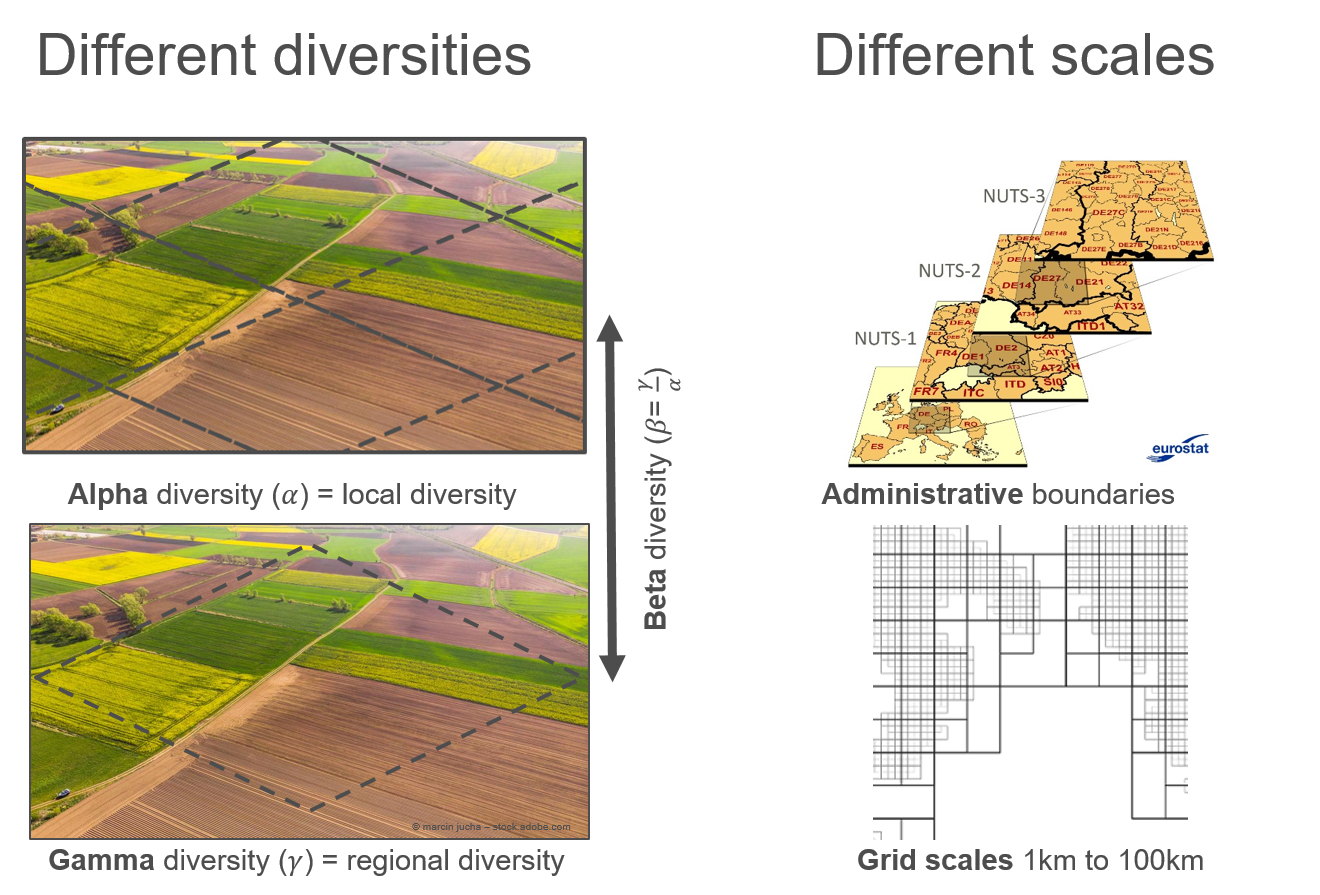} \vspace*{-.5cm}
    \caption{Definitions of α-, β- and γ-diversities and visualisation of different spatial scales — administrative regions or grids at different resolutions — adopted in this study. \textbf{α-diversity}: Denotes local diversity and is computed at the finest scale of 1 km x 1 km.
\textbf{γ-diversity}: Represents regional diversity, computed at various administrative regional aggregations (from national to sub-national levels) or in coarser resolution grids ranging from 2 km x 2 km to 100 km x 100 km.
\textbf{β-diversity}: Defined as the ratio between γ and α, it serves as a measure of inter-ecosystems diversity.}
    \label{fig:Fig-A}
\end{figure*}

For the analysis of diversity dependency on the spatial domains, we compute the diversity using a set of spatial resolutions that are consistent with the EUROSTAT reference grids (i.e. 1 km, 2 km, 5 km, 10 km, 50 km and 100 km)\footnote{The grids used for this study were obtained from the EUROSTAT GISCO platform and are available at \url{https://ec.europa.eu/eurostat/web/gisco/geodata/reference-data/grids}.}. These nested grids are widely used by Member States (MS) \footnote{Countries that are part of the European Union are referred to as Member States (e.g. in 2018 the United Kingdom was a Member State).} which facilitates the use of our results for policy relevant reporting. 
We also consider spatial aggregations corresponding to EU administrative units NUTS (Nomenclature des Unités territoriales statistiques) at the national (NUTS 0) and subnational (NUTS 2) level. This analysis is particularly useful since these administrative units are used for statistical reporting and thus directly relevant for policy evaluation and implementation. The sizes of the administrative units vary from country to country. To calculate crop diversity at the administrative levels, all 1 km grid cells located on administrative boundaries are discarded. This pragmatic selection avoids attribution to multiple NUTS regions. The area contribution of these grid cells, quantified in \ref{tab:coverage_border_cells_N0} for all countries, are relatively small, with less than 5\% of  all 1 km cells discarded  for all countries except for Luxembourg and Slovenia (14.2\% and 6.2\% respectively). \ref{fig:Fig-A} illustrates graphically the geographic units of observation used in this study.

\subsection{Crop diversity} 
\label{subsec:crop_div_methodo}

Shannon entropy is a widely used index of diversity \citep{uthes2020farm,mahy2015simulating,egli2021crop,weigel2018crop} that can be applied to assess crop diversity at different spatial and temporal scales \citep{schaak2023long}. The entropy value rather gives uncertainty in the species identity of a sample, than the diversity of the species in the community \citep{jost2006entropy}. The diversity index, calculated as an exponent of the Shannon entropy \citep{shannon1948mathematical}, creates equivalence classes among communities, converting entropy into "effective number of species" \citep{macarthur1965patterns}. The computation of the diversity at different spatial scales, follows the approach  of \cite{jost2007partitioning,tuomisto2010diversity,aramburu2020scale}. The literature often defines the scale-dependency of crop diversity using the terms α, β, γ, as depicted in the conceptual framework (see \ref{fig:Fig-A}).  α-diversity corresponds to local diversity while the  γ-diversity corresponds to crop diversities computed at larger spatial (regional) scales. The β-diversity is the ratio between γ-diversity and α-diversity. While α-diversity can be considered representative of the rotations occurring at the level of large farms or a cluster of several smaller farms, β-diversity is more linked to the differences of cropping systems between adjacent regions \citep{aramburu2020scale}. We adopt a finest reference scale at 1 km resolution to compute α-diversity. With $M$ the number of  1 km cells in the domain (one grid cell or one NUTS) on which the diversity is computed, with $S$ the number of crop types considered and with $c_{ij}$ the count of 10 meters pixels of the EU crop map for the crop type $j$ in the 1 km cell $i$, we compute the α-diversity, γ-diversity, β-diversity respectively in equations  \ref{eq:alpha-div}, \ref{eq:gamma-div}, \ref{eq:beta-div} as:

\begin{equation}
    \alpha= \exp \Bigg( - \sum_{i=1}^{M} \sum_{j=1}^{S}\frac{c_{ij}}{\sum_{k=1}^{M} \sum_{l=1}^{S} c_{kl}} \ln \bigg(\frac{c_{ij}}{ \sum_{l=1}^{S} c_{il} }\bigg) \Bigg)
    \label{eq:alpha-div}
\end{equation}

\begin{equation}
    \gamma= \exp \Bigg( - \sum_{j=1}^{S}\frac{\sum_{i=1}^{M} c_{ij}}{\sum_{k=1}^{M} \sum_{l=1}^{S} c_{kl}} \ln \bigg(\frac{\sum_{i=1}^{M} c_{ij}}{ \sum_{k=1}^{M} \sum_{l=1}^{S} c_{kl} }\bigg) \Bigg)
    \label{eq:gamma-div}
\end{equation}

\begin{equation}
    \beta = \frac{\gamma}{\alpha}
    \label{eq:beta-div}
\end{equation}

The demonstration coming from the equations (17a) and (17b) of \citet{jost2007partitioning} is found in \ref{eq:alpha-div-sup} and \ref{eq:gamma-div-sup} .

\begin{figure*}[h!!]
    \centering
    
        \centering
        \includegraphics[width=0.9\textwidth]{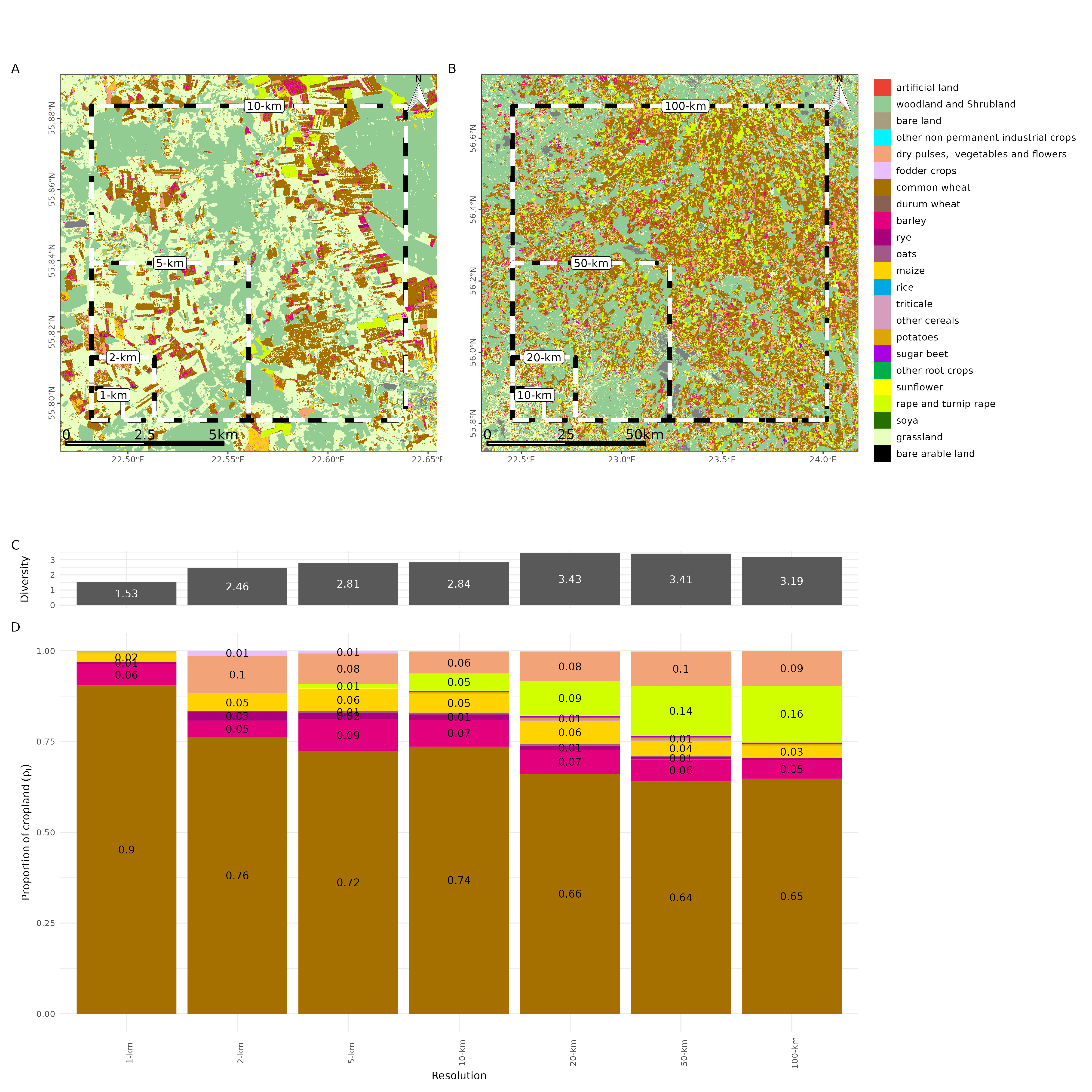}

    \caption{Crop diversity is computed at different grid scales ranging from 1 to 100 km as illustrated here for a region in Latvia-Lithuania. For each subfigure, panel A shows the scales 1-km, 2-km, 5-km and 10-km with the crop map in background. Panel B shows the scales 10-km, 20-km, 50-km and 100-km with the crop map in background. Panel C shows the γ-diversity for the respective sample squares shown in A and B. Panel D shows the proportion of crop types for the different scales. These proportions are used to compute the Shannon diversity. }
    \label{fig:Fig-B}
\end{figure*}

These conceptual diversity definitions translate into the different diversity indicators that are computed in this study. We compute the diversity at different scales (grids and administrative) and evaluate changes at subsequent aggregation levels. We consider the 1 km scale as a representation of local farm level diversity, the 2 km to 5 km scale to describe landscapes, and 10 km to 100 km representing landscape to regional or national levels depending on the size of the country. Note that α-diversity computed for a 1 km grid, $\alpha_{1km}$, would correspond to $\gamma_{1km}$, being the γ-diversity computed for a 1 km grid. The diversity D (either $\alpha$, $\beta$ or $\gamma$) can be computed at larger spatial scales either considering grids at coarser resolution (from $D_{2km}$ to $D_{100km}$ for 2 km to 100 km, respectively) or administrative aggregations ($D_{NUTS_{0}}$, $D_{NUTS_{2}}$ at national and sub-national scales, respectively). Note that the α-diversity computed at a coarser resolution grid than 1 km, would still come from the reference scale. For illustration, with the sub-grid 10 meter EU crop map in background,  \ref{fig:Fig-B} A and B show the grids cells corresponding to the 1, 2, 5, 10 km and 10, 20, 50, 100 km scales, respectively, for four contrasted regions (Spain, Northern Italy, and Latvia-Lithuania). For the grid cell in \ref{fig:Fig-B} A and B, the panels C and D show the explicit crop distribution and the corresponding γ-diversity values at each scale (see also in \ref{fig:Fig-B-examnded} for similar figures  in different contrasted regions).


Aggregated statistics of α- and γ-diversity from subnational to national level can provide useful information for e.g. the CAP agricultural sector resilience monitoring framework \citep{cmef}. The notation used in this study for the aggregation of diversity D (either $\alpha$, $\beta$ or $\gamma$) are $\overline{D}_{NUTS_0}$, $\overline{D}_{NUTS_2}$ and $\overline{D}_{R}$ respectively referring to the crop diversity computed at grid scale and averaged by subnational, national levels or whatever region $R$.

EUROSTAT, FAO (Food and Agriculture Organisation) or country reported crop statistics at national and regional level can in principle be used to respectively compute ${D}_{NUTS_0}$ and  ${D}_{NUTS_2}$ (if the reporting is based on the same crops listed in \ref{tab:eucropmap_class} and reported values are complete). However, the diversity at regular grid resolution can only be computed from higher-spatial resolution data such as the ones derived from Earth Observation data we employed.

\subsection{Summarizing crop diversity computed across scales}
\label{sec:method_sum_crop_div_across_scales}

In order to account for all the scales of observation of crop diversity observed at country level and for all EU-28 countries, we compute:
\begin{itemize}
    \item $avg(\gamma)$ the average across scales of the γ-diversity computed at grid scale 
    \item $std(\gamma)$ the standard deviation across scales of the γ-diversity computed at grid scale
\end{itemize}

We then compute the difference of those entities between the country level and the EU-28 level, noted as $\Delta avg(\gamma)$ for the average and $\Delta std(\gamma)$ for the standard deviation.

\subsection{Farm Structure Survey}
\label{sec:nethod_fss}

The EUROSTAT Farm Structure Survey{\interfootnotelinepenalty10000 \footnote{See \url{https://ec.europa.eu/eurostat/statistics-explained/index.php?title=Glossary:Farm_structure_survey_(FSS)}}} (FSS) provides comprehensive information on farm operations in the agricultural sector in the EU. The 2016 FSS collected data samples on the characteristics of agricultural holdings, including their location, size, production, crops grown, livestock, and farming practices. Here, we use the 2016 FSS to calculate the average farm size at the subnational and national levels, to assess whether there is a relationship with the  α-diversity (local) in the different administrative regions. We further discuss the link to the size of the farm and the limitations and perspectives of using such statistical surveys (section \ref{sec:discussion}).

\subsection{Crop diversity in the USA}
\label{sec:method_usa}

Our study follows a similar approach as \citet{aramburu2020scale} in the USA, with a difference in the chosen (finest) reference resolution for $\alpha$ (1 km) which is smaller and corresponds to 100 ha instead of the 392 ha for the study in the USA. Apart from our alignment with a grid system pre-defined for policy analysis, our choice also considers the fact that European farms are on average smaller than US farms, as we discuss later in detail (see Section \ref{sec:discussion}). We compare our resulting α, β, γ diversities computed at grid scale in the EU with the results obtained by \citet{aramburu2020scale} in the USA for the conterminous United States (2008-2017).

\section{Results} 
\label{sec:results}

This section presents a comprehensive analysis of crop diversity across EU-28 countries, using various aggregation levels. We start by comparing local α, β and γ-diversity at administrative national and regional scales (section \ref{subsec:alpha_gamma_nuts}). We then cross-examine crop diversity behaviour at different grid scales and build a typology for all EU-28 countries (section \ref{subsec:alpha_gamma_grid}). We also inspect the relationship between local diversity and farm size data from the FSS conducted in 2016 (section \ref{subsec:fss}) and finally we compare our results with those obtained in the USA by \citet{aramburu2020scale} \ref{subsec:res-usa-eu}.

\begin{figure}[h]
    \centering 
    \includegraphics[width=0.48\textwidth,height=0.55\textwidth]{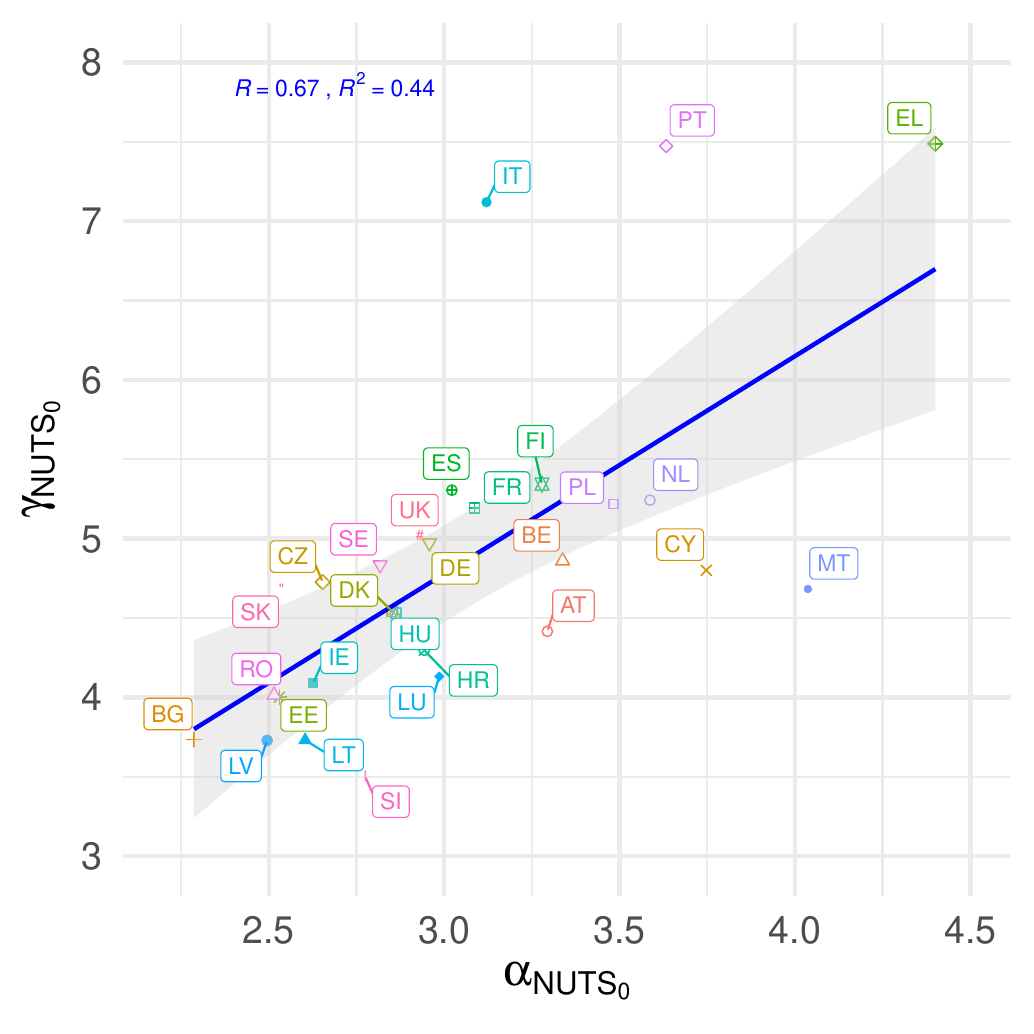} \vspace*{-.5cm}
    \caption{Fit of α-diversity ($\alpha_{NUTS_{0}}$) against γ-diversity ($\gamma_{NUTS_{0}}$) at national level (see country abbreviations in \ref{tab:countrycodes}).}
    \label{fig:Fig-Ka}
\end{figure} 

\subsection{Comparing α, β and γ diversity at national and regional scale}
\label{subsec:alpha_gamma_nuts}

\begin{figure*}[!h]
    \centering 
        \includegraphics[width=\linewidth]{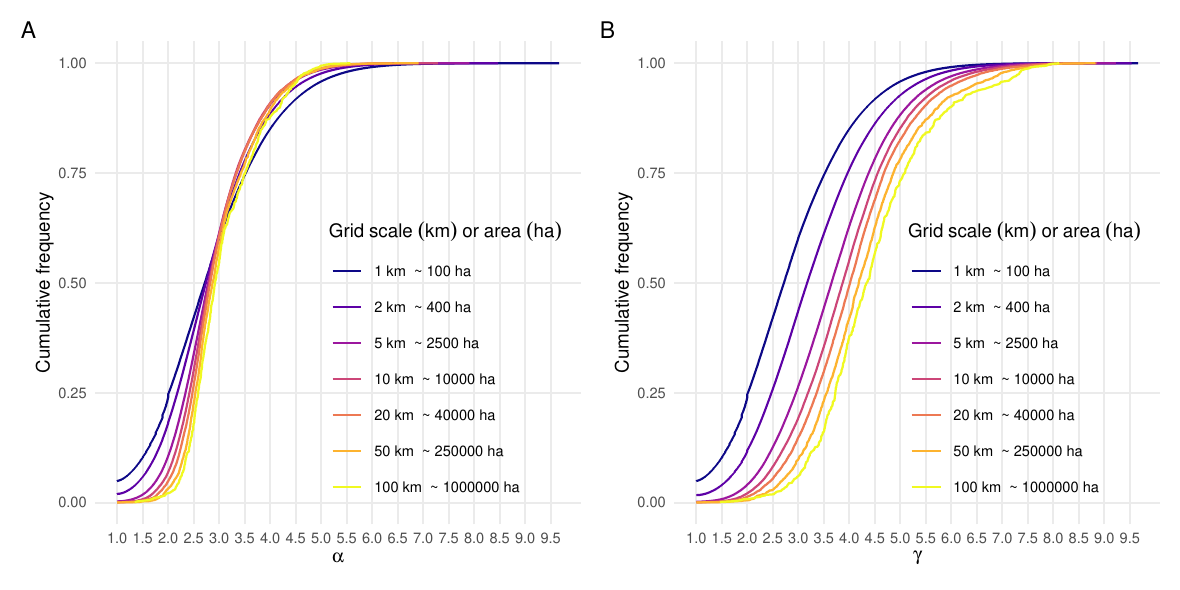}
    \caption{Cumulative distribution of α-diversity (A) and γ-diversities (B) for  1 km to 100 km by grid scale in the EU-28. On figure (B), we see that there is probability of $50\%$ that  $\alpha_{1km}=\gamma_{1km}$ takes a value of less or equal to 2.55 and that $\gamma_{100km}$ takes a value of less or equal to 4.3.}
    \label{fig:Fig-C}
\end{figure*}

We compare α-diversity and  γ-diversity at national  (see \ref{fig:Fig-Ka}) and subnational (see \ref{fig:Fig-Kb-bis}) levels. These results are also mapped in EU-28 for α, γ and β diversity respectively in \ref{fig:Fig-E}, \ref{fig:Fig-H} and \ref{fig:Fig-Kb}. We find a large variation across countries for $\alpha_{NUTS_{0}}$ (range 2.3-4.4), $\gamma_{NUTS_{0}}$ (range 3.5-7.5) and $\beta_{NUTS_{0}}$  (range 1.2-2.3). Lowest α-diversity is obtained for Bulgaria ($\alpha_{BG}=2.3$), lowest γ-diversity is obtained for Slovenia ($\gamma_{SI}=3.5$), which is a relatively small country with uniformy wet climate, and lowest β-diversity is obtained for Malta ($\beta_{MT}=1.2$), probably because of size contraints as well. Greece shows the highest  α-diversity and γ-diversity ($\alpha_{EL}=4.4$ and $\gamma_{EL}=7.5$) that are probably linked to the beneficial effects on the Mediterranean climate. We observe that for most of the countries in EU-28, with low α-diversity ($\alpha_{NUTS_{0}}< 3.5$), the β-diversity varies little ($\beta_{NUTS_{0}}$ between 1.5 and 1.7), regardless whether the analysis was performed at national or subnational level (see \ref{fig:Fig-Kb-bis}). This explains the moderate to strong linear relationship between  α-diversity and  γ-diversity ($R=0.67$ at national and $R=0.81$ at subnational levels). Most of the countries analysed therefore show a country-level crop diversity driven by a spatial variability around 1.5 times greater in space than the local farm level. For countries with a high local crop diversity ($\alpha_{NUTS_{0}}> 3$), we distinguish two extreme opposite behaviours. On the one hand,  Malta (MT) and Cyprus (CY) show α-diversity almost identical to the national level γ-diversity ($\beta_{NUTS_{0}} < 1.3$), again due to the fact that they are both relatively small countries.  On the other hand, Portugal (PT) and Italy (IT) are at least twice more diverse at the country level than at 1 km (α) scale (highest $\beta_{NUTS_{0}} >2$ as seen in \ref{fig:Fig-Kb}). 
Although Portugal (PT) and the Netherlands (NL) have similar local diversities ($\alpha_{PT}=3.6$ and $\alpha_{NL}=3.5$), their national-scale diversities are remarkably different ($\gamma_{PT}=7.5$ and $\gamma_{NL}=5.2$), with the high value for Portugal showing the heterogeneity of crop production across the country. The high local  diversity in the Netherlands is primarily driven by the diversity in the Eastern part of the Netherlands, where the majority of cropland is found (e.g. in Flevoland NL23). Slovakia seems to grow at local scale a modest number of crop types ($\alpha_{SK} = 2.5$) while at the country level a higher number of crop types are grown $\gamma_{SK} = 4.7$, resulting in a high $\beta_{SK} = 1.85$. We conclude that no obvious generic relation can be found between α (farms-level) diversity and regional or national diversity as probably the different cultures and the complex historical heritage in Europe also contributed in shaping the diversity of agricultural practices that are found nowadays.

\subsection{Quantifying crop diversity across scales of observation}

\label{subsec:alpha_gamma_grid}

\begin{figure*}[h]
    \centering 
    \includegraphics[width=0.97\textwidth,height=0.50\textwidth]{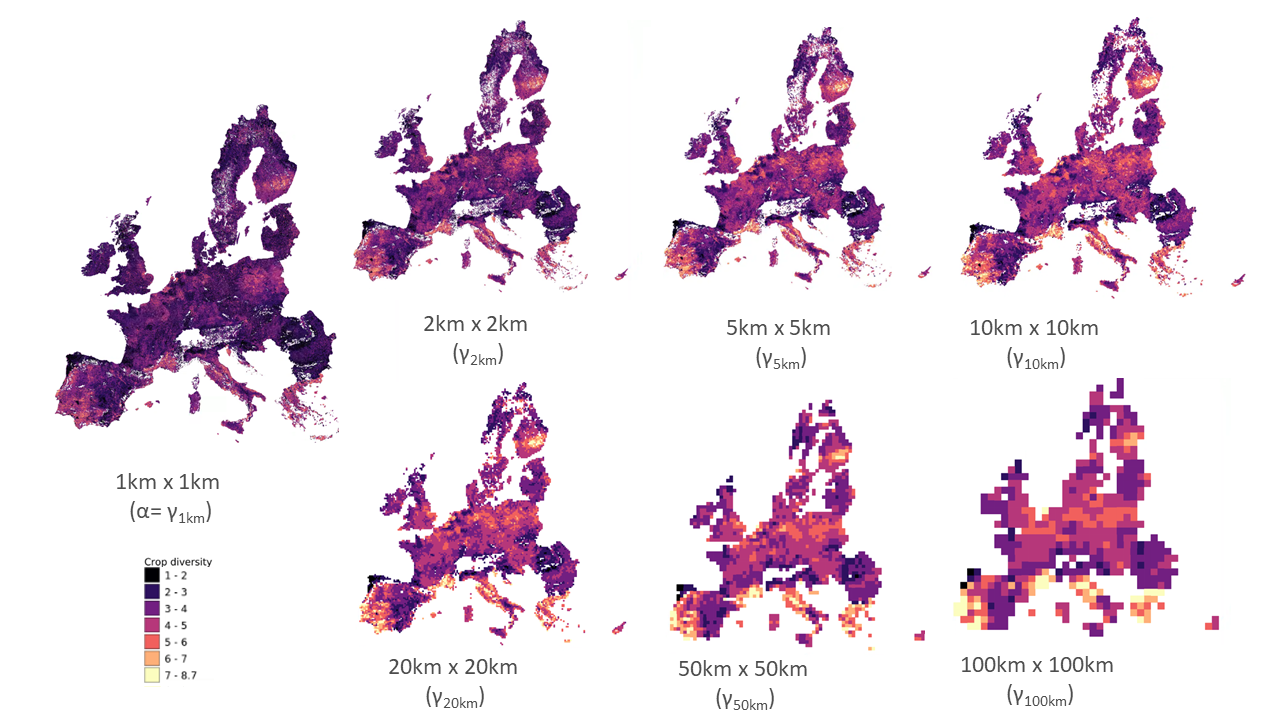} 
    \caption{Maps of the EU-28 with γ-diversity computed for all grid scales (1 km, 2 km, 5 km, 10 km, 50 km and 100 km). Grid cells with less than 1\% of cropland area have been filtered out in order to remove the effect of erratic 10 m x 10 m pixels from the EU crop map. This threshold is arbitrary and can be tweaked by any data user (see Section \ref{sec:code_and_data}).}
    \label{fig:Fig-D-map}
\end{figure*} 
\begin{figure*}[h]
    \centering 
    \includegraphics[width=.85\textwidth]{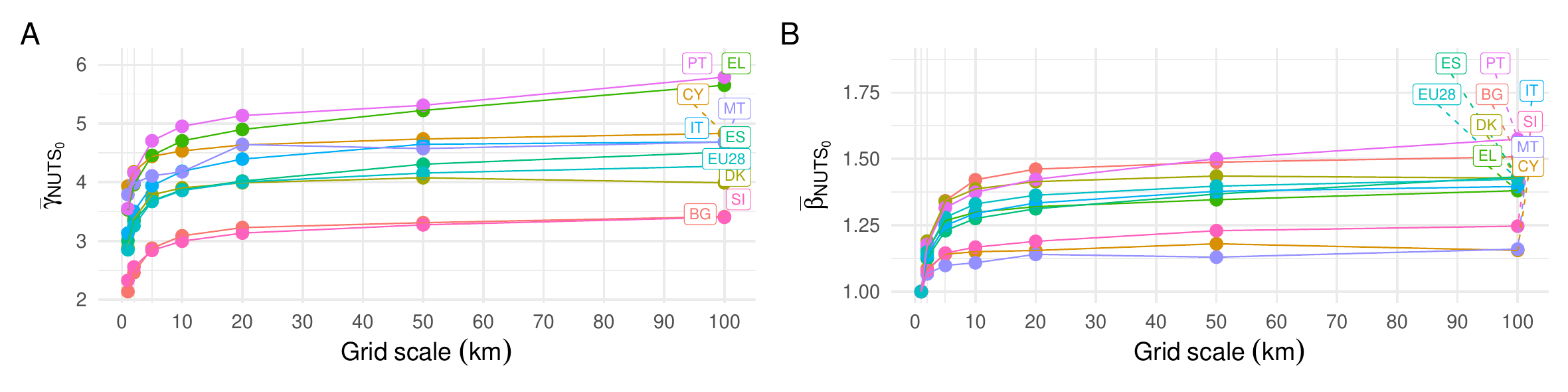} \vspace*{-.5cm}
    \caption{ γ-diversity and β-diversity for each scale, averaged by country of the EU-28 (respectively $\overline{\gamma}_{{NUTS}_0}$ and $\overline{\beta}_{{NUTS}_0}$).}
    \label{fig:Fig-D}
\end{figure*}

\begin{figure*}[h]
    \centering 
    \includegraphics[width=.90\textwidth]{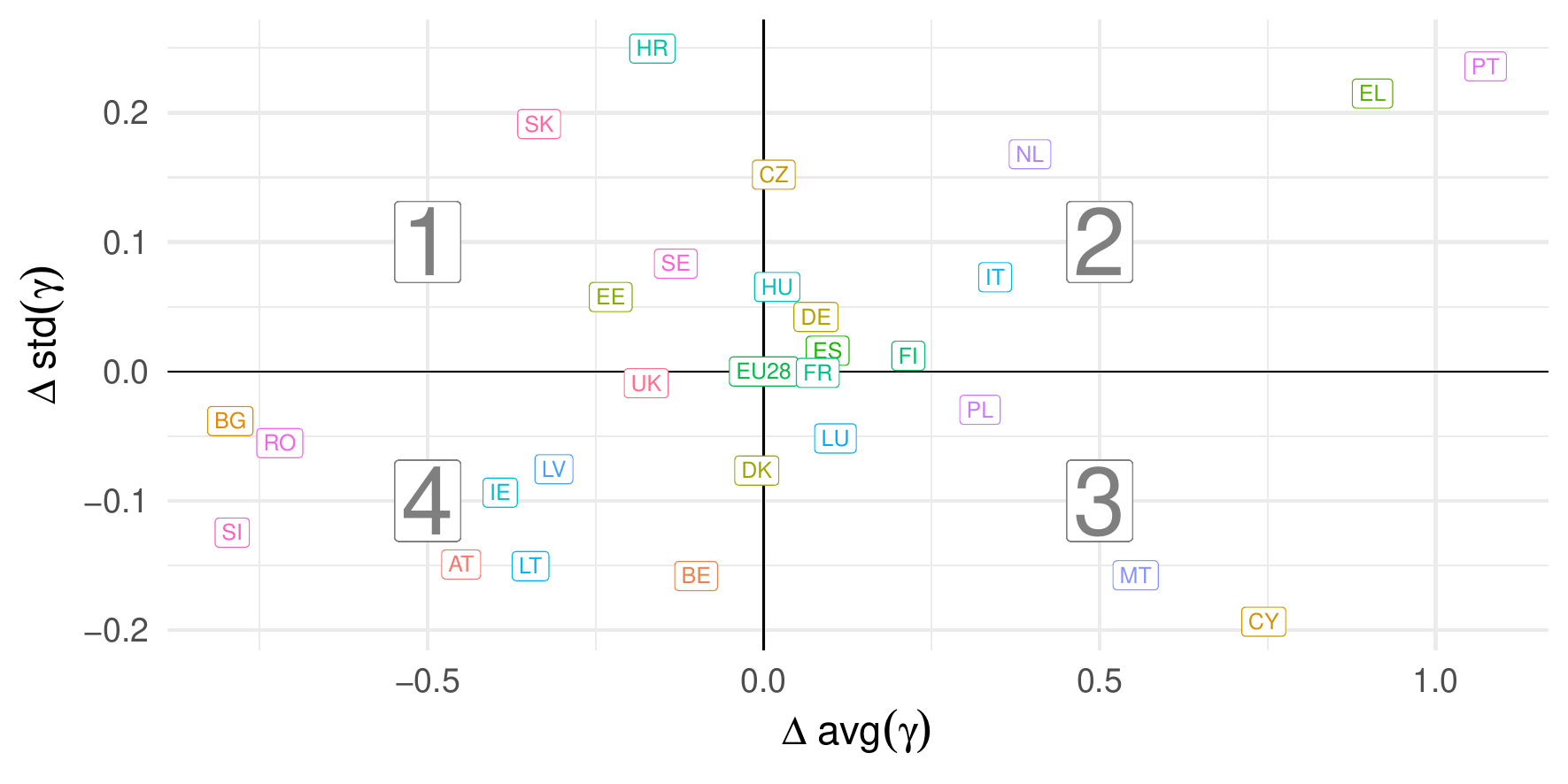} 
    \vspace*{-.5cm}
    \caption{Clustering of countries based on the difference with the EU-28 along the x-axis and y-axis, respectively for the average (avg) and standard deviation (std) of γ-diversity over scales.}
    \label{fig:Fig-D-scatter-plot-delta-div}
\end{figure*}

Geometric regions are more suited for characterizing the impact of analysis scale on crop diversity than administrative regions, which may differ greatly in size,  pedoclimatic, socio-economic and landscape typologies. We compare the diversity values obtained at different scales for the entire EU-28, and then present and discuss the results aggregated at national and sub-national administrative levels. 

The exploration of the crop diversity scale dependency is conducted through the analysis of the cumulative distributions of the α-diversity and  γ-diversity computed at all grid scales. \ref{fig:Fig-C} (A) demonstrates that the overall median α-diversity is not affected by the computational scale. Conversely, the frequency of occurrence of low and high α-diversity values is gradually reduced at the larger scales. This is expected because, at larger scales, the tails of the distribution have larger probabilities to be mixed with more common α-diversity values, so the occurrence of the extreme values is largely reduced. \ref{fig:Fig-C} (B) reveals a clear tendency of increasing γ-diversity at larger scales. The distributions rapidly shift from local diversity ($\gamma_{1km}=\alpha_{1km}$), with an EU-28 average of 2.85, with a sharp increase to higher scales diversity up to 3.86 at 10 km ($\gamma_{10km}$) and remain steadier up to 4.27 at 100 km ($\gamma_{100km}$). The cumulative distribution highlights that at the 1 km scale, 60 \% of EU-28 cropland has less than 3 equally abundant crops ($\gamma_{1km}<=3$). However, using the 100 km scale, up to 95\% of Europe's agricultural area produces more than 3 (equivalent) crops ($\gamma_{100km}>=3$). 

Mapping crop diversities at different scales for all EU-28 countries (\ref{fig:Fig-D-map}) shows that crop diversity, across all scales, is greater in southern Europe (in particular the Mediterranean) than in northern Europe. In these Mediterranean regions, favourable agro-climates in valleys and plains sustain heterogeneous cultivation,  whilst mountainous regions (the Alps, the Pyrenees, Western Border of Sweden) and regions with a majority of rangeland (west-Normandie in France, Galicia in Spain, Ireland) obtain lower values of crop diversity.

\ref{fig:Fig-D} display the γ-diversity (A) and β-diversity (B) as a function of grid scales (in $km$) for a sample of countries extracted from our analysis in \ref{fig:Fig-Ka} in section \ref{subsec:alpha_gamma_nuts} for the EU-28. Whether for countries with high β-diversity (Italy (IT), Portugal (PT)), low  β-diversity (Malta (MT), Cyprus (CY), Slovenia (SI)) or with similar β-diversity but with low (Bulgaria (BG)), medium (Denmark (DK), Spain(ES)) or high (Greece (EL)) α-diversity, we note an apparent logarithmic behaviour for each country. \ref{fig:Fig-D} reveals that diversity at higher scales is strongly co-determined by the diversity at the lowest scales. This finding corroborates the importance of implementating local level farm management strategies for national level benefits. The comparison between the γ-diversity (\ref{fig:Fig-D} (A)) and β-diversity (\ref{fig:Fig-D} (B)) highlights interesting features of the across-scale diversity dependency. While different aspects are driving the observed variability of γ-diversity, the normalization operation to compute the β-diversity (i.e. equation \ref{eq:beta-div}) filters out the role of the α-diversity. This allows us to better highlight the differences between the α- and γ-diversity through a condensed measure of the diversity-scales relationship. Two main classification types can arise from the visual inspection of \ref{fig:Fig-D} (B). First, the size of the relative change from small to large scales allow classifying different countries. Countries like Malta (MT), Cyprus (CY) and Slovenia (SI) display small relative changes. Countries like Bulgaria (BG), Italy (IT), Portugal (PT) and Denmark (DK) show the larger relative changes. Secondly, the β-diversity inspection allows to better quantify whether there is a saturation scale where the diversity stops growing (e.g. Bulgaria, Denmark, Cyprus) or, instead, if it continues to grow also at larger scales, as it happens in the majority of the larger countries.   Moreover, the proportional increase of  γ-diversity over scales emphasizes the importance of national-scale strategies to encourage the growth of more crop types from landscape to national scale. Based on the previous findings, we can characterise the crop diversity across scales for each country by 1) the magnitude of  γ-diversity and 2) the relative increase of  γ-diversity along scales. \ref{fig:Fig-D-scatter-plot-delta-div} displays this interaction by plotting the average of the γ-diversity versus the standard deviation of the γ-diversity with their respective EU-28 value subtracted to normalize the comparison respectively $\Delta avg(\gamma)$  and $\Delta std(\gamma)$. This characterisation, enables the classification of the countries in four groups, represented in each quadrant (number indicated in grey in the graph). Compared to EU28, in:
\begin{itemize}
    \item Quadrant (1) the country has a lower diversity and the diversity is less uniform across scales;
    \item Quadrant (2) the country has a higher diversity, less uniform across scales;
    \item Quadrant (3) the country has a higher diversity, but more uniform across scales;
    \item Quadrant (4) the country has a lower diversity, but more uniform across scales.
\end{itemize}

\subsection{Link with average farm size from the European Farm Structure Survey}
\label{subsec:fss}

\begin{figure*}[!t]
    \centering 
    \includegraphics[width=.85\textwidth]{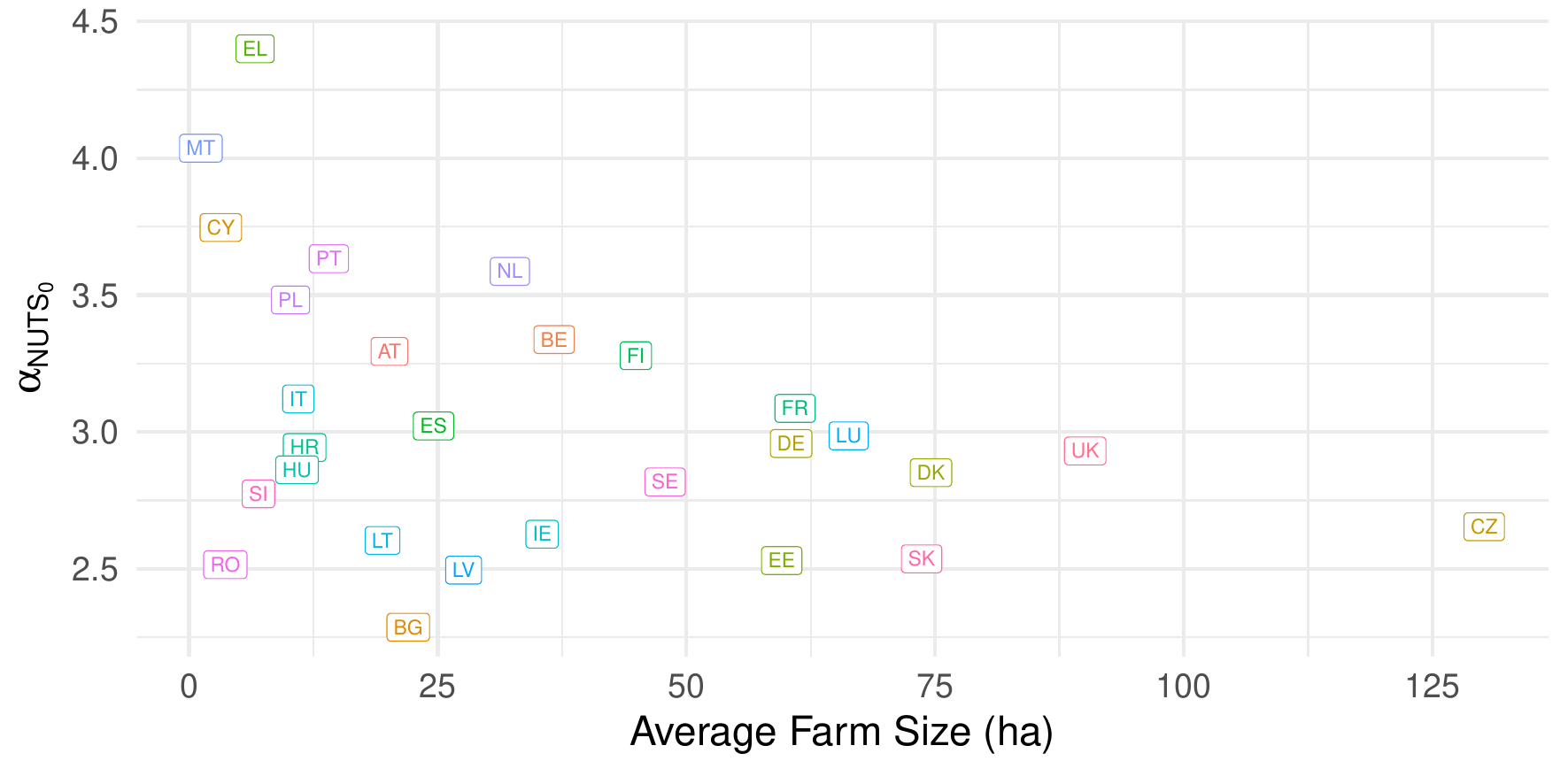} 
    \caption{α-diversity as a function of average farm size (in hectares) at national level in EU-28.
}\vspace*{-.5cm}
    \label{fig:Fig-N-fss-diversity-n0}
\end{figure*}

In 2016, the EU-28 had 10,467,760 farmers cultivating a total of 173,338,550 ha of Utilised Agricultural Area (UAA), corresponding to an  EU average farm size of 16.56 ha with an important disparity across the continent (\ref{tab:fss_nuts0}). In this section, we examine the potential relationships between farm size and α-diversity. \ref{fig:Fig-N-fss-diversity-n0} and \ref{fig:Fig-N-fss-diversity-n2}  respectively display  the farm size against $\alpha_{NUTS_{0}}$ and  $\alpha_{NUTS_{2}}$. We find that the highest local crop diversity values ($\alpha_{NUTS_{0}}>3.7$)  are only found in countries with very small average farm sizes (less than 10 ha), e.g. in  Malta (MT), Cyprus (CY) and Greece (EL). Malta and Cyprus are noteworthy islands and among the smallest countries in the EU-28. We do not imply that farms smaller than 10 ha are needed to reach very high levels of local crop diversity. Evidence shows that countries with the largest average farm size (over 50 ha) are associated with relatively lower local crop diversity ($\alpha_{NUTS_{0}}<3.1$). Examples of such countries include United Kingdom (UK), Slovakia (SK), Denmark (DK), Germany (DE), France (FR) and Luxembourg (LU) presenting medium values of local crop diversity ($2.1<\alpha_{NUTS_{0}}<3.1$). The Czech Republic (CZ), which boasts the largest farm size in Europe, exceeding 150 ha on average, also shows one of the lowest values ($\alpha_{NUTS_{0}}=2.7$). Although the hypothesis that small farm size is concomitant to high local crop diversity may appear attractive, we have identified  contrasting instances in the cases of Romania (RO) and Slovenia (SI) where farm size $< 10$ ha coincides with $\alpha_{NUTS_{0}}<2.7$. Of all analysed countries, Bulgaria (BG) has the lowest α-diversity ($\alpha_{BG}=2.3$) despite having a relatively low average farm size of almost 25 ha. 
Interestingly, the Netherlands (NL) shows at sub-national level (\ref{fig:Fig-N-fss-diversity-n2}) disparities with the highest values of mean local crop diversity ($\alpha_{NUTS_{2}}>4.3$) for the Zeeland (NL34) and  Flevoland (NL23) administrative regions and average farm sizes around 50 ha, but also regions with low $\alpha_{NUTS_{2}}$ around $2.1$ and average farm size around 40 ha. In some regions in Spain (ES), despite a relatively small average farm size around 20 ha, also low  $\alpha_{NUTS_{2}}$ around 2 are found,  e.g. in Asturias (ES12), Cantabria (ES13) and Galicia (ES11). Besides agriculture, these regions are highly covered by meadows and forests. In summary, there is no conclusive relation between farm sizes and local crop diversity in EU-28. However, some robust  patterns appear, such as no high local diversity is achieved in regions dominated by large farm sizes.

\subsection{Comparing crop diversity in the EU and in the USA}
\label{subsec:res-usa-eu}

\begin{figure*}[!t]
    \centering 
    \includegraphics[width=0.85\textwidth,height=0.45\textwidth]{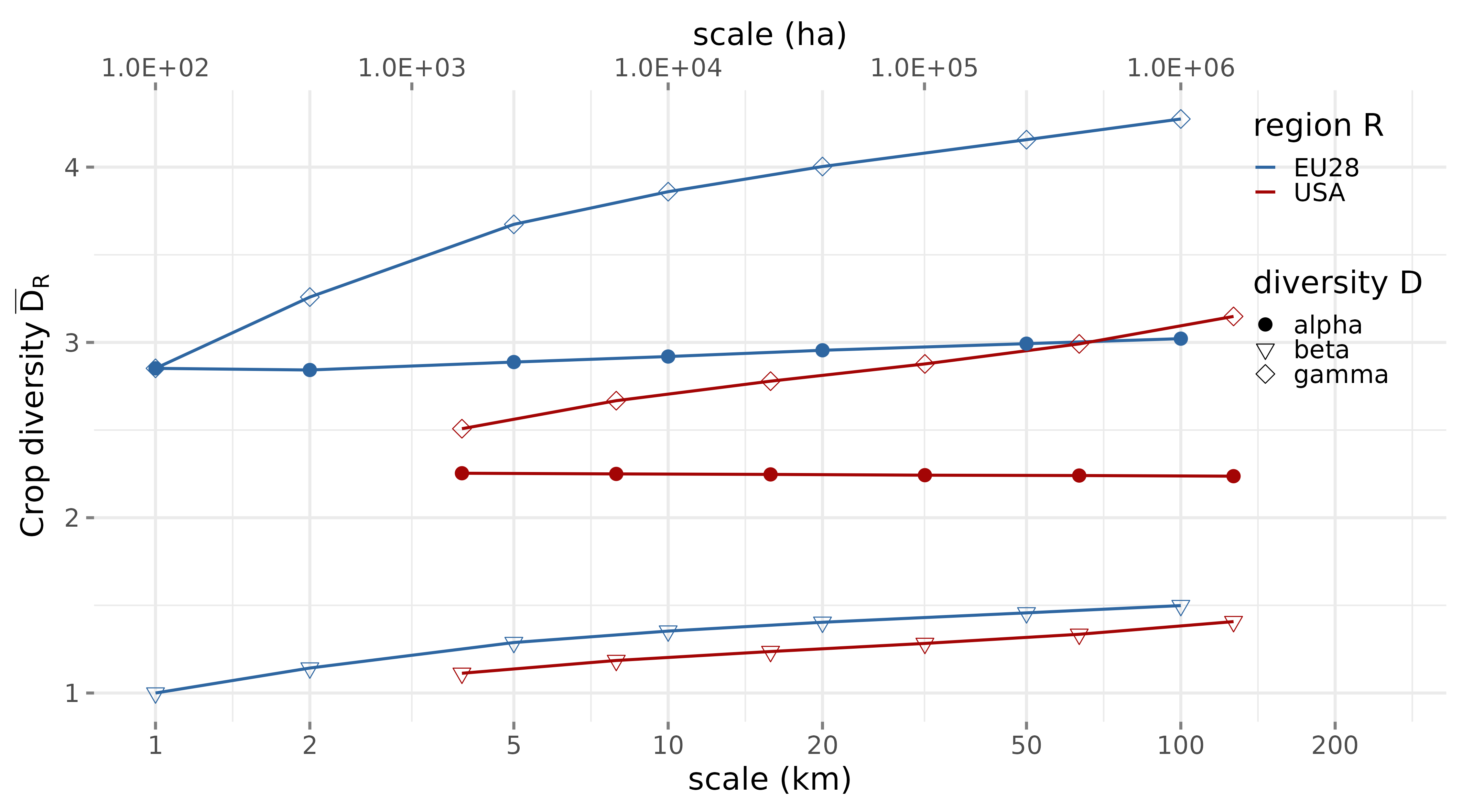} \vspace*{-.5cm}
     \caption{Comparison of α, β, γ diversity computed across scales averaged for EU-28 countries (2018) in our study and averaged for the conterminous United States (2008-2017) in the study of \citet{aramburu2020scale}. The graph is in semi-logarithmic scale (for the x-axis). }
    \label{fig:Fig-D-pnas-merlot2020-comparison_2}
\end{figure*}

All EU-28 countries exhibit relatively high crop diversity (α, β, γ) across scales compared to the United States, as shown in \ref{fig:Fig-D-pnas-merlot2020-comparison_2}. We observe that γ-diversity is higher in the EU-28 than in the USA. The growth of γ-diversity is more linear for the USA than for the EU-28, disclosing a logarithmic growth (as already pointed out in section \ref{subsec:alpha_gamma_nuts}). This finding reveals that the increase of γ-diversity with scale of analysis is $40-50\%$ greater in the EU-28 than in the USA. The α-diversity is higher in EU-28 (close to 3) than in the USA (slightly over 2), despite being measured on a smaller grid of scale of 100 ha for EU-28 and 392 ha for USA. If both regions were compared using the same grid size, the difference would be higher as diversity increases with grid size (as shown in section \ref{subsec:alpha_gamma_grid})

\section{Discussion and outlook} 
\label{sec:discussion}


We use a novel EO-derived crop map dataset to compute a comprehensive continuous mapping of crop diversity in 28 EU countries. The dataset allows enhanced spatial analysis compared to survey based derived from a limited number of farms. We demonstrate that local (1 km) crop diversity, in the EU representative for large farms, or clusters of small-to-medium sized farms,  is not linked in a homogeneous way (i.e., a unique β-diversity across scales and countries) with regional and national crop diversity. For example, we find a near-linear behaviour for countries with low to medium local crop diversity ($\alpha_{NUTS_{0}}<3.5$). For small countries like Malta or Cyprus  α-diversity is almost identical to the national level γ-diversity. Other factors limiting an increase of gamma diversity with scale include forests dominant landscapes (e.g. Sweden) and limited variability in pedoclimatic conditions (e.g. Belgium). EU policies  are usually devised, implemented and monitored at subnational (NUTS 2) or national (NUTS 0) subnational levels. By exploring the β-diversity from local to regional or national level, we are able to evaluate how crop diversity increases from farm level to the larger  administrative levels.    The varying levels of β-diversity indicate the necessity for specific policy consideration when targeting crop diversity across European countries.

In this work, we computed the crop diversity at the standard grids scales endorsed by EUROSTAT, to facilitate a possible linkage of our analysis to  data used by countries to report on their strategies. We deliver for the first time a continuous mapping of the EU at 1, 2, 5, 10, 20, 100 km of crop diversity and reveal a logarithmic shape increase of γ-diversity and β-diversity with scale. Since the average farm size in the EU spans from  1.21 ha (Malta) to 130.25 ha (Czechia), future work  could consider introducing computations at 100 m and 500 m grid scale (respectively 1 ha and 25 ha) allowing a better estimation of farm-level diversity across EU countries. We note however, given the 10 m resolution of the EU crop map, that the crop diversity computed at 1 ha would at best integrate 100 grid points, adding a challenge for reaching statistical significance.

In \ref{fig:Fig-D-scatter-plot-delta-div}, we classify crop diversity (γ-diversity) in individual countries according to magnitude and slope of changes, and place them in 4 quadrants characterising response typologies for policies, relative to the other countries that include crop diversity:

\begin{itemize}
\item Quadrant (1) improving the country crop diversity would need a country wide approach, with more attention drawn to specific regions;
\item Quadrant (2) further improving the country diversity would mostly need addressing specific regions;

\item Quadrant (3) the country already has good crop diversity. Further improvements could be achieved by addressing diversification across scales;
\item Quadrant (4) policies improving crop diversity across all scales would be most useful.
\end{itemize}

For instance, the top two food producers and economies in EU-28, Germany (DE) and France (FR), are part of quadrant (2). In these countries, depending on the policy objective, a diversification strategy could specifically address regions with low farm diversity (e.g. Centre-Val de Loire region in France or Bavaria in Germany). Romania (RO), Slovenia (SI) and Bulgaria (BG),  part of quadrant 4, could benefit from improvement of crop diversity across scales. Indeed the national scale crop type distribution  in \ref{fig:Fig-D-GammaGridValues} panel B, shows that in these countries 85\% of the total cropland represented by only three crop types. Refinement of these typologies should be further elaborated considering other agri-environmental factors for transversal  analysis. A policy strategy can include the evaluation of the ecological services provided by crop diversity at scales that could go even below the farm size. Future studies can be devoted to evaluate the appropriate scale chosen for α-diversity with respect to different biological functions such as farmland birds, pollinators' presence, or grassland butterflies related indicators. 

While large γ-diversity over larger scales or administrative regions is relevant for long-term production stability of countries, large local-scale diversity is important for the resilience of agro-ecosystems and of the individual farms. Our analysis suggests three interesting patterns linking farm size and farm level crop diversity in the EU-28. We find that high local diversity (α-diversity larger than 3.7) is achieved in countries with generally small average farm sizes (less than 10 ha) as observed in Greece, Malta and Cyprus. In contrast, small values of local diversity (α-diversity smaller than 2.7) are also found in countries with relatively small average farm size (less than 10 ha) like in Romania or Slovenia. With the largest average farm sizes (more than 75 ha), Czech Republic, United Kingdom, Denmark and Slovakia  have relatively low values of local crop diversity (less than 3.1). We note here, that there is a great potential in enhancing the EO based analysis of crop diversity and related outcomes with data from statistical surveys  such as  the 2020 FSS{\interfootnotelinepenalty10000 \footnote{carried every ten years. See https://ec.europa.eu/eurostat/web/agriculture/census-2020}, the Farm Accountancy/Sustainability Data Network (FADN/FSDN)}{\interfootnotelinepenalty10000 \footnote{COM/2022/296 final, \url{https://eur-lex.europa.eu/legal-content/EN/TXT/?uri=COM:2022:296:FIN}}}  or data collected from the farmer’s applications for subsidies through the IACS with declarations done at holding level (Van der Velde, 2024). Importantly, working with these data sources will allow characterizing crop diversity at parcel and holding level. Further exploring this synergy should lead to better policy anticipation, implementation and evaluation.


We show that the crop diversity is higher in the EU-28 that in the USA. The growth with increasing scale of γ-diversity is more linear for the USA than for the EU-28, the latter disclosing a logarithmic growth. In the case of EU-28, country sizes span from 1e4 ha to 1e7 ha which is comparable to the range of states' sizes in the USA.  In the case of the USA, an exponential growth of the γ-diversity is noticed for scales greater than 1e7 hectares (their computation goes up to 1e9 ha). Computing the crop diversity at higher scales would be interesting to see if EU-28 follows a similar trend for scales encompassing several countries.

We acknowledge the inherent limitations of our study, particularly in the selection of crop classes for the EU crop map, which may not fully capture certain crops and the ecosystem services delivered. However, this study's primary focus is to introduce an initial understanding of crop diversity across the EU, using the current classifications available. Therefore, our analysis lays the groundwork for future research to explore functional diversity between crop species, as suggested by \cite{di2010agrobiodiversity} and applied in Sweden by \cite{schaak2023long}. Their approach underlines that crop diversity assessment is not solely a matter of scale but also requires thematic considerations, relying on different or more nuanced groupings among crop types to better represent ecosystem services. 

Furthermore, \cite{aramburu2020scale} show how multi-annual analysis of crop type maps delivers insights into temporal developments in crop diversity. While for this study, focusing on the EU28, multiple years were not available , this situation will evolve soon. In particular, the European Environment Agency (EEA) has foreseen to deliver annual time series of crop maps for the  38 countries at 10m resolution from 2017 onwards\footnote{The Crop Types map are part of the Copernicus HRL-VLCC (High Resolution Layer - Vegetation Land Cover Component).}. However, since in that product, the selection of crop types is different from the EU crop map used in this publication, crop diversity maps and calculations will probably somewhat differ.

Finally, our crop diversity scale analysis could provide important information for analysis of farm and farm sector resilience across scales, and enrich our understanding the connection of resilience and ecological benefits associated with cropping systems. This information is equally relevant from a scientific and policy perspective.

\section{Conclusions}

Our study underscores the critical role of spatial scales in understanding crop diversity in the EU-28, using a novel high-resolution EO derived crop map. We explore the scale dependency of crop diversity,  from farm to landscape, regional and national scales (using grid scales from 1 km to 100 km). We also compute crop diversity at subnational (NUTS 2) and national (NUTS 0) administrative levels, usually required for EU policy monitoring and reporting.  Our research highlights the disparities between regions and countries in terms of α, β, and γ diversity, emphasizing the importance of regional specificity in crop cultivation. Our findings confirm that country level crop diversity (γ-diversity) has a complex relationship with farm level diversity (α-diversity), especially for countries with high values of α-diversity. We investigate a characterisation of the crop diversity across scales, finding a logarithmic shape of dependency for γ-diversity and β-diversity. We propose a classification of countries into four groups, depending on the magnitude and disparity of the γ-diversity across scales, enabling generalized policy-relevant typologies for each of those groups. With the aim  of linking to farm-level information, we reveal that very high local crop diversity (more than 3.7) is only achieved in the EU-28 countries with small farm sizes (less than 10 ha) such as in Greece, Malta and Cyprus (nothing that the two latest are islands and among the smallest countries in the EU-28). In addition, we show that crop diversity in the EU-28 shows higher values than in the USA. 

These EO-based crop diversity calculations will feed in the further development of performance monitoring indicators of the CAP, as well as the regional targeting of practices that stimulate more diverse crop-mixes through the CAP implementation in the EU. Looking forward, the availability of annual crop type maps will enable temporal comparisons, and the exploration of ecosystem co-variates from various sources will deepen our understanding of the link to ecosystem services. This research underscores the need for continued investigation into the complex interplay between crop diversity and agricultural resilience, with a particular focus on the role of spatial scale.

\section*{Code and Data}
\label{sec:code_and_data}
The code, primarily written in Python, PostGIS, and R, is publically available \citep{JRC129557}. The computational burden is about 3 days to compute yearly indicators. The processing was performed on the JRC Big \ac{BDAP} platform \citep{soille2018versatile}. Therefore the code can be readily used to provide regular (annual) updates in order to monitor crop diversity changes.

In addition to the code, some raster datasets have been produced \citep{JRC129557} and are also available. Based on the 10-m resolution EU crop map 2018, the following metrics for scales 1, 2, 5, 10, 20, 50, 100 km have been derived:
\begin{itemize}

\item \textit{crop diversity} (α, β, γ) and \textit{crop richness}, for each scale (see Figure 1 for overview)\footnote{\url{https://jeodpp.jrc.ec.europa.eu/ftp/jrc-opendata/DRLL/CropDivV1.1/2018/CropDiversity/}}
\item  \textit{crop type proportion}, for each crop and each scale \textit{over cropland}\footnote{\url{https://jeodpp.jrc.ec.europa.eu/ftp/jrc-opendata/DRLL/CropDivV1.1/2018/CropProportionCropland/}}
\item  \textit{crop type proportion}, for each crop and each scale \textit{over all land}\footnote{\url{https://jeodpp.jrc.ec.europa.eu/ftp/jrc-opendata/DRLL/CropDivV1.1/2018/CropProportionLand/}}
\end{itemize}

We also provide for the \textit{crop diversity} (α, β, γ) at subnational (NUTS 2)  and national (NUTS 0) levels \citep{JRC129557} in tables \footnote{\url{https://jeodpp.jrc.ec.europa.eu/ftp/jrc-opendata/DRLL/CropDivV1.1/2018/CropDiversityByNuts/}}.

As illustration of the different data made available along with the manuscript, the reader can inspect the proportion of each crop at 1km (\ref{fig:supp_crop_percent_1km}) and the crop richness at different scales 
(\ref{fig:supp_crop_richness_allscales}). Although some of the aforementioned datasets have not been directly used to compute the presented results, they may be inspirational for future studies.

Finally, we provide a viewer \footnote{\url{https://jeodpp.jrc.ec.europa.eu/eu/dashboard/voila/tree/REFOCUS/CropDiversity2018V1.1/CropDiversity.ipynb} available upon sign up} to access at grid and administrative scale the α, β, γ diversity, granting for advanced analysis depending on users' needs. We will maintain the viewer access and maintenance as long as possible.

\section*{Author contributions}
Conceptualization, M.M, M.Z and R.D.; Data curation, M.M. and R.D.; Investigation, M.M., M.Z.and R.D.; Methodology,  M.C., M.M., M.Z and R.D.; Supervision, F.D. and R.D.; Validation, M.M.; Visualization, M.M. and R.D.; Writing—original draft, M.M., M.Z and R.D.; Writing—Review and editing, F.D., M.C., M.M., M.V, M.Z. and R.D. All authors have read and agreed to the published version of the manuscript.

\section*{Acknowledgements}
We thank the authors of \cite{aramburu2020scale} for providing the machine-readable results data of their publication behind some of the figures. We also thank Davide de Marchi, Data Governance and Services - Joint Research Centre, for developing the viewer of the data related to this publication.

\bibliography{sample.bib}

\appendix

\onecolumn

\clearpage

\setlength{\cftfignumwidth}{1.4cm}
\setlength{\cfttabnumwidth}{1.5cm}

\tableofcontents
\listoffigures
\listoftables

\setcounter{figure}{0}
\setcounter{table}{0}
\setcounter{page}{1}
\clearpage
\section*{Supplementary material}

\label{AppendixA}
\captionsetup{list=no}
\renewcommand{\thetable}{Supplementary Table \arabic{table}}

\begin{table}[h]
\caption{The classified crops in the EU crop map along with the corresponding code classes.}
\label{tab:eucropmap_class}
\centering
\begin{tabular}{@{}ll@{}}
\toprule
Class & Label                                       \\ \midrule
211   & Common wheat                                \\
212   & Durum wheat                                 \\
213   & Barley                                      \\
214   & Rye                                         \\
215   & Oats                                        \\
216   & Maize                                       \\
217   & Rice                                        \\
218   & Triticale                                   \\
219   & Other cereals                               \\
221   & Potatoes                                    \\
222   & Sugar beet                                  \\
223   & Other root crops                            \\
230   & Other non permanent industrial crops        \\
231   & Sunflower                                   \\
232   & Rape and turnip rape                        \\
233   & Soya                                        \\
240   & Dry pulses, vegetables and flowers          \\
250   & Other fodder crops (excl. temp. grasslands) \\
290   & Bare arable land                            \\ \bottomrule
\end{tabular}
\end{table}

\begin{table}[]
\caption{Average farms size by country in 2016 (source EUROSTAT FSS).}
\label{tab:fss_nuts0}
\centering
\begin{tabular}{@{}lrrr@{}}
\toprule
Country        & Farm - number & Utilised agricultural area - hectare & Average size \\ \midrule
Belgium        & 36,890        & 1,354,250                            & 36.71        \\
Bulgaria       & 202,720       & 4,468,500                            & 22.04        \\
Czechia        & 26,530        & 3,455,410                            & 130.25       \\
Denmark        & 35,050        & 2,614,600                            & 74.60        \\
Germany        & 276,120       & 16,715,320                           & 60.54        \\
Estonia        & 16,700        & 995,100                              & 59.59        \\
Ireland        & 137,560       & 4,883,650                            & 35.50        \\
Greece         & 684,950       & 4,553,830                            & 6.65         \\
Spain          & 945,020       & 23,229,750                           & 24.58        \\
France         & 456,520       & 27,814,160                           & 60.93        \\
Croatia        & 134,460       & 1,562,980                            & 11.62        \\
Italy          & 1,145,710     & 12,598,160                           & 11.00        \\
Cyprus         & 34,940        & 111,930                              & 3.20         \\
Latvia         & 69,930        & 1,930,880                            & 27.61        \\
Lithuania      & 150,320       & 2,924,600                            & 19.46        \\
Luxembourg     & 1,970         & 130,650                              & 66.32        \\
Hungary        & 430,000       & 4,670,560                            & 10.86        \\
Malta          & 9,210         & 11,120                               & 1.21         \\
Netherlands    & 55,680        & 1,796,260                            & 32.26        \\
Austria        & 132,500       & 2,669,750                            & 20.15        \\
Poland         & 1,410,700     & 14,405,650                           & 10.21        \\
Portugal       & 258,980       & 3,641,690                            & 14.06        \\
Romania        & 3,422,030     & 12,502,540                           & 3.65         \\
Slovenia       & 69,900        & 488,400                              & 6.99         \\
Slovakia       & 25,660        & 1,889,820                            & 73.65        \\
Finland        & 49,710        & 2,233,080                            & 44.92        \\
Sweden         & 62,940        & 3,012,640                            & 47.87        \\
United Kingdom & 185,060       & 16,673,270                           & 90.10        \\
\hline \\
EU-28          & 10,467,760    & 173,338,550                          & 16.56        \\ \bottomrule
\end{tabular}%
\end{table}

\begin{table}[!h]
    \centering
    \caption{Proportion of grid cells including NUTS 0 border cells (in \%)}
    \label{tab:coverage_border_cells_N0}
    \begin{tabular}{cccccccccccccc}
        \toprule
         AT & BE & BG & CY & CZ & DE &  DK & EE & EL & ES & FI   & FR & HR & HU \\ \midrule
         3 & \cellcolor[HTML]{FFBE00} 4.9 & 1.1 & 0 & 2.8 & 1 & 0.2 & 0.8 & 0.3 & 0.4 & 0.2 &0.3 & 2.1 & 2.3 \\ \midrule \midrule
           IE & IT & LT & LU & LV & MT & NL & PL & PT & RO & SE & SI & SK & UK   \\ \midrule
          0.8 & 0.2 & 1.2 & \cellcolor[HTML]{FFBE00}14.2 & 1.6 & 0 & 3.3 & 0.6 & 1.7 & 0.6 & 0.2 & 
           6.2 & 3.2 & 0.2  \\
        \bottomrule
    \end{tabular}%
\end{table}

\begin{table}[h]
\centering
\caption{List of Country Codes and Names}
\label{tab:countrycodes}
\begin{tabular}{@{}ll@{}}
\toprule
\textbf{Country Code} & \textbf{Country Name} \\ \midrule
AT & Austria \\
BE & Belgium \\
BG & Bulgaria \\
CY & Cyprus \\ 
CZ & Czech Republic \\ 
DE & Germany \\ 
DK & Denmark \\ 
EE & Estonia \\ 
EL & Greece \\ 
ES & Spain \\ 
FI & Finland \\
FR & France \\ 
HR & Croatia \\
HU & Hungary \\ 
IE & Ireland \\ 
IT & Italy \\ \
LT & Lithuania \\ 
LU & Luxembourg \\
LV & Latvia \\ 
MT & Malta \\ 
NL & Netherlands \\ 
PL & Poland \\ 
PT & Portugal \\ 
RO & Romania \\ 
SE & Sweden \\ 
SI & Slovenia \\ 
SK & Slovakia \\
UK & United Kingdom \\ 
\hline\\
\end{tabular}
\end{table}
\renewcommand{\theequation}{Supplementary Eq. S\arabic{equation}}

We define $M$ the number of  1 km cells in the domain (one grid cell or one NUTS) on which the diversity is computed, $S$ the number of crop types considered and $c_{ij}$ the count of 10 meters pixels of the EU crop map for the crop type $j$ in the 1 km cell $i$. We also define $p_{ij}$ as the proportion of 10 meters pixels count of the EU crop map for the crop type $j$ over the cropland 10 meters pixels count in the 1 km cell $i$  as in \ref{eq:pij}:

\begin{equation}
    p_{ij}=\frac{c_{ij}}{\sum_{l=1}^{S}c_{il}}
        \label{eq:pij}
\end{equation}
We then specify $w_{i}$ the weight applied on cell $i$ and defined as the proportion of cropland 10 meters pixels count of the EU crop map in the 1 km cell $i$  over the total cropland 10 meters pixels count in the domain of interest, as in \ref{eq:wi}:
\begin{equation}
    w_{ij}=\frac{\sum_{l=1}^{S}c_{il}}{\sum_{k=1}^{M} \sum_{l=1}^{S} c_{kl}}
        \label{eq:wi}
\end{equation}

We finally compute the γ-diversity and  α-diversity respectively from Equation (17a) and (17b) of \citet{jost2007partitioning} in \ref{eq:alpha-div-sup} and \ref{eq:gamma-div-sup}:
 
\begin{align}
   \alpha &=\exp \Bigg(- \sum_{i=1}^{M} w_i \sum_{j=1}^{S} p_{ij} \ln p_{ij} \Bigg) \nonumber\\
   &= \exp \Bigg(- \sum_{i=1}^{M} \bigg( \frac{\sum_{l=1}^{S}c_{il}}{\sum_{k=1}^{M} \sum_{l=1}^{S} c_{kl}} \bigg) \sum_{j=1}^{S}  \bigg( \frac{c_{ij}}{\sum_{l=1}^{S}c_{il}} \bigg) \ln \bigg( \frac{c_{ij}}{\sum_{l=1}^{S}c_{il}} \bigg)\Bigg) \nonumber\\
   &= \exp \Bigg(- \sum_{i=1}^{M} \sum_{j=1}^{S} \bigg( \frac{\sum_{l=1}^{S}c_{il}}{\sum_{k=1}^{M} \sum_{l=1}^{S} c_{kl}} \frac{c_{ij}}{\sum_{l=1}^{S}c_{il}} \bigg) \ln \bigg( \frac{c_{ij}}{\sum_{l=1}^{S}c_{il}} \bigg) \Bigg) \nonumber\\
  &= \exp \Bigg( - \sum_{i=1}^{M} \sum_{j=1}^{S}\frac{c_{ij}}{\sum_{k=1}^{M} \sum_{l=1}^{S} c_{kl}} \ln \bigg(\frac{c_{ij}}{ \sum_{l=1}^{S} c_{il} }\bigg) \Bigg)\label{eq:alpha-div-sup}   
\end{align}

\begin{align}
    \gamma &= \exp \Bigg(  \sum_{j=1}^{S} \bigg( -\sum_{i=1}^{M} w_i p_{ij}\bigg) \ln \bigg( \sum_{i=1}^{M} w_i p_{ij} \bigg)\Bigg)\nonumber \\
    &= \exp  \Bigg( \sum_{j=1}^{S} \bigg( -\sum_{i=1}^{M} \Big( \frac{\sum_{l=1}^{S}c_{il}}{\sum_{k=1}^{M} \sum_{l=1}^{S} c_{kl}} \frac{c_{ij}}{\sum_{l=1}^{S}c_{il}} \Big)\bigg) \ln \bigg( \frac{\sum_{l=1}^{S}c_{il}}{\sum_{k=1}^{M} \sum_{l=1}^{S} c_{kl}} \frac{c_{ij}}{\sum_{l=1}^{S}c_{il}}\bigg)\Bigg) \nonumber \\
    &=\exp \Bigg( - \sum_{j=1}^{S}\frac{\sum_{i=1}^{M} c_{ij}}{\sum_{k=1}^{M} \sum_{l=1}^{S} c_{kl}} \ln \bigg(\frac{\sum_{i=1}^{M} c_{ij}}{ \sum_{k=1}^{M} \sum_{l=1}^{S} c_{kl} }\bigg) \Bigg)
    \label{eq:gamma-div-sup}
\end{align}
\renewcommand{\thefigure}{Supplementary Fig. S\arabic{figure}}

\begin{figure*}[]
    \centering 
    \includegraphics[width=1\textwidth]{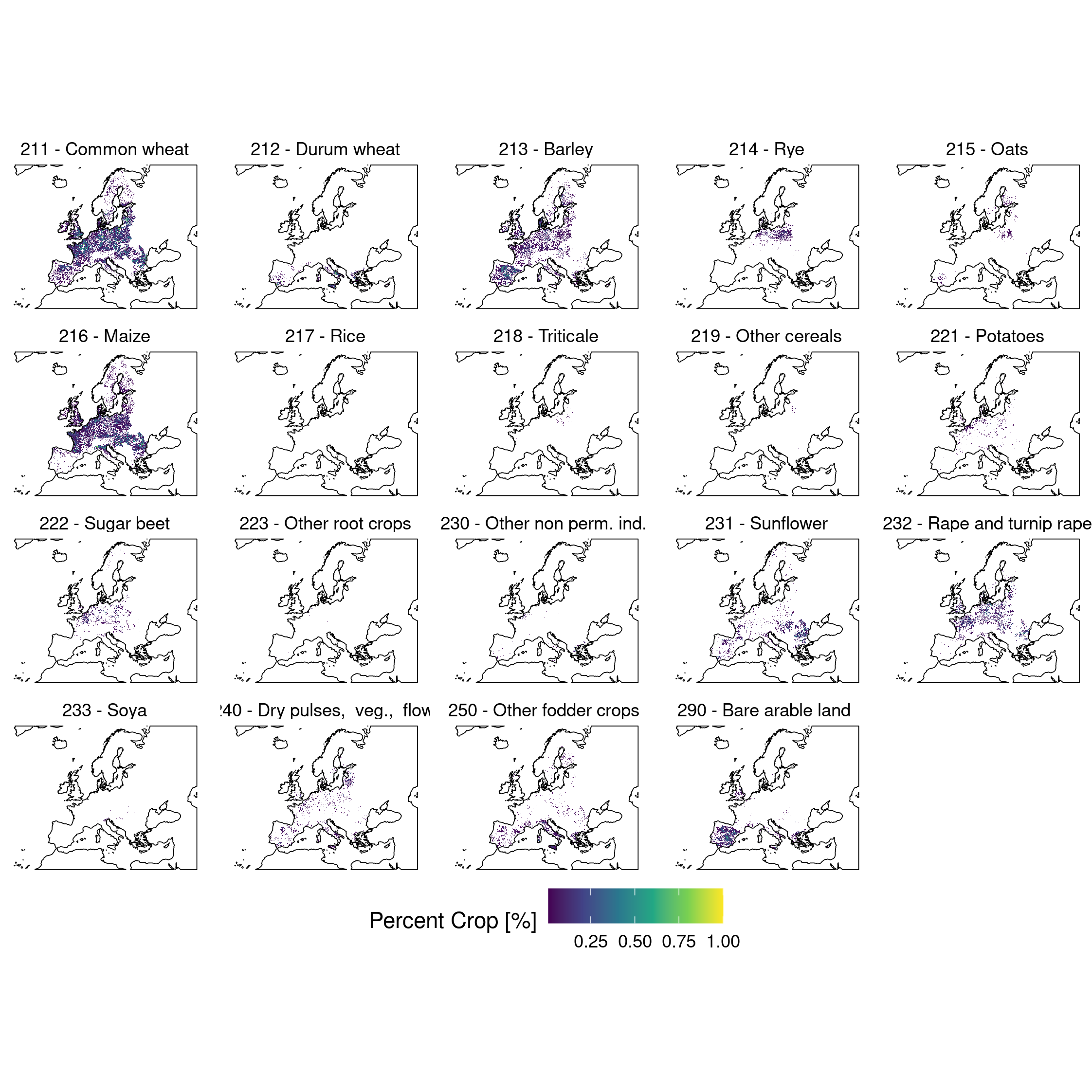} \vspace*{-.5cm}
    \caption{Cropland coverage for different crop types at 1 km scale}
    \label{fig:supp_crop_percent_1km}
\end{figure*} 

\begin{figure*}[]
    \centering 
    \includegraphics[width=0.71\textwidth]{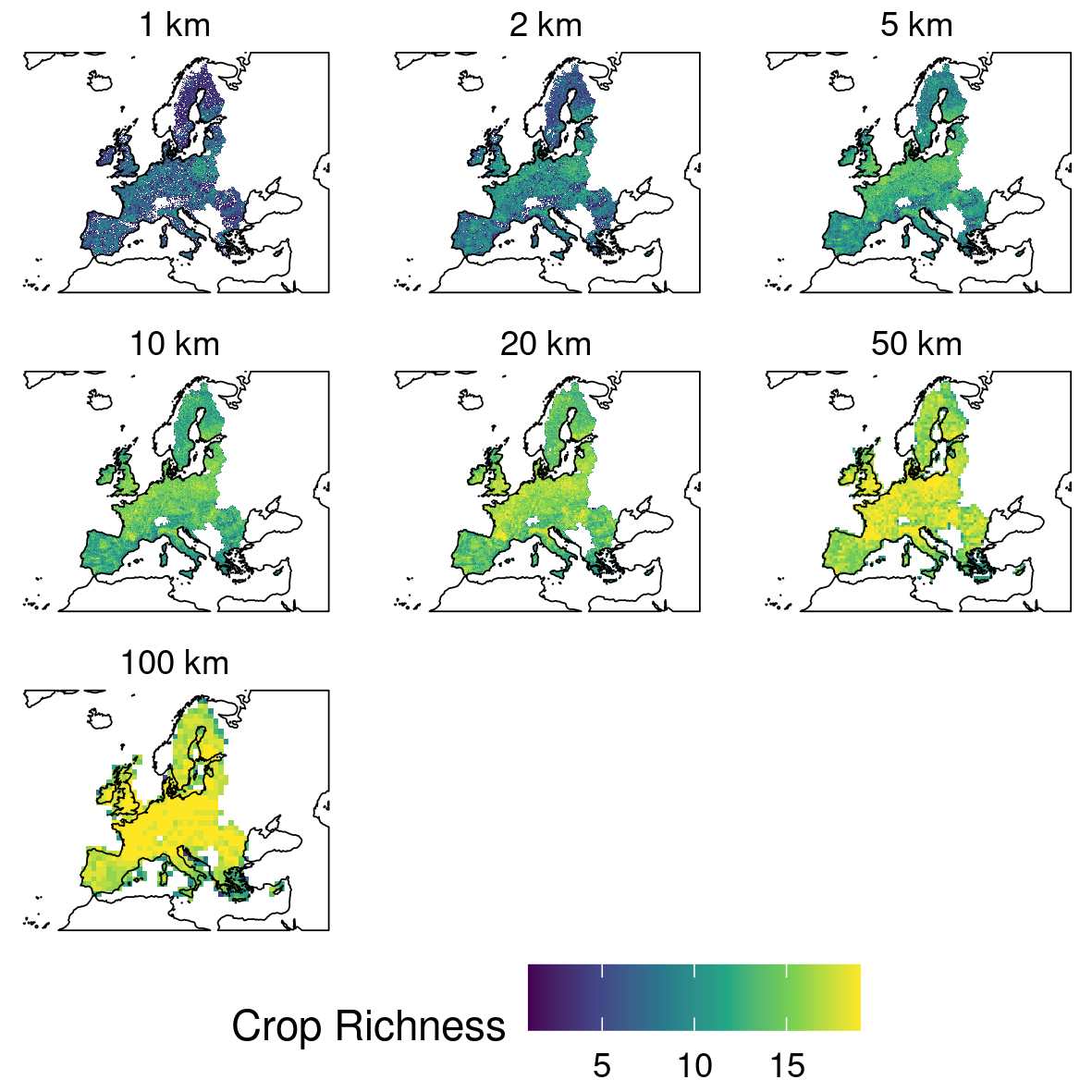} \vspace*{-.5cm}
    \caption{Crop richness at different scales.}
    \label{fig:supp_crop_richness_allscales}
\end{figure*} 

\begin{figure*}[h!!]
    \centering
    
    \begin{subfigure}[b]{\textwidth}
        \centering
        \includegraphics[width=0.9\textwidth]{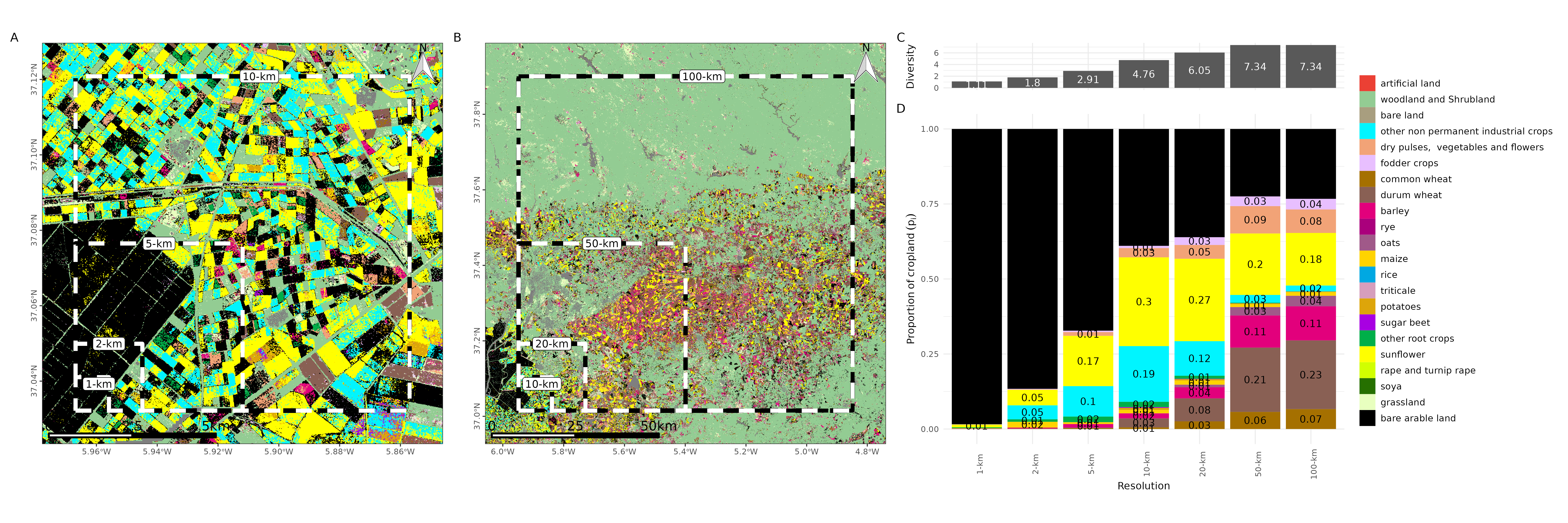}
        \caption{Spain (N1700000E2900000)}
        \label{fig:subfig1}
    \end{subfigure}
    
    \begin{subfigure}[b]{\textwidth}
        \centering
        \includegraphics[width=0.9\textwidth]{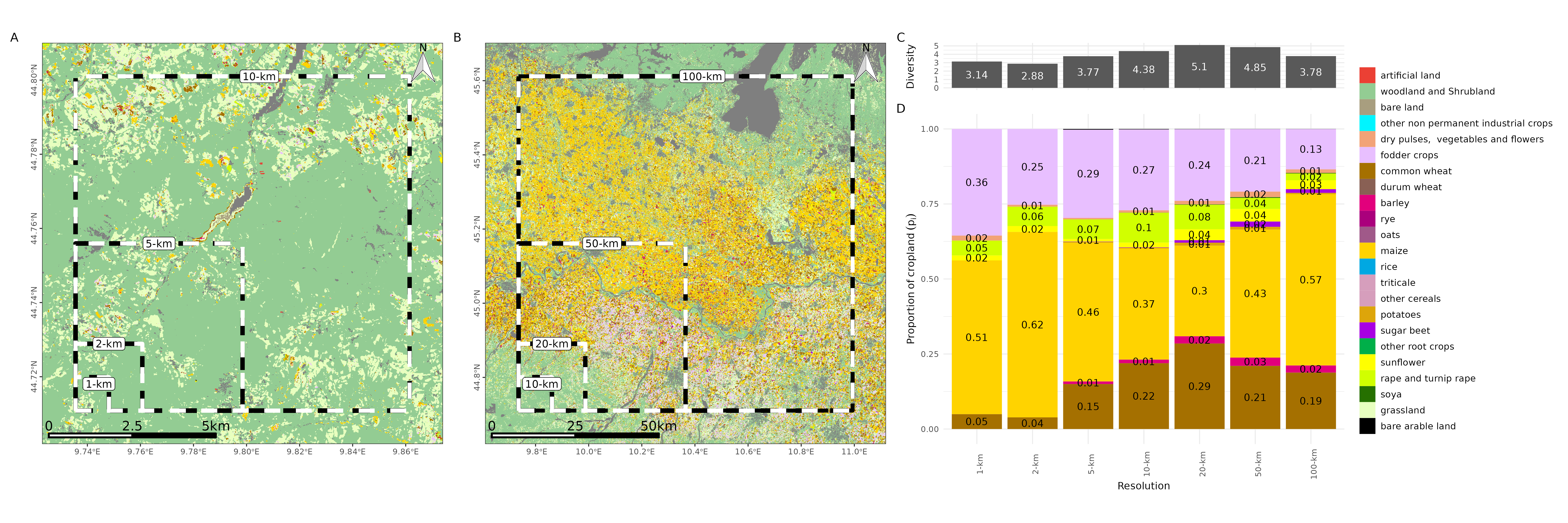}
        \caption{Northern Italy (N2400000E4300000 )}
        \label{fig:subfig2}
    \end{subfigure}
    
    \begin{subfigure}[b]{\textwidth}
        \centering
        \includegraphics[width=0.9\textwidth]{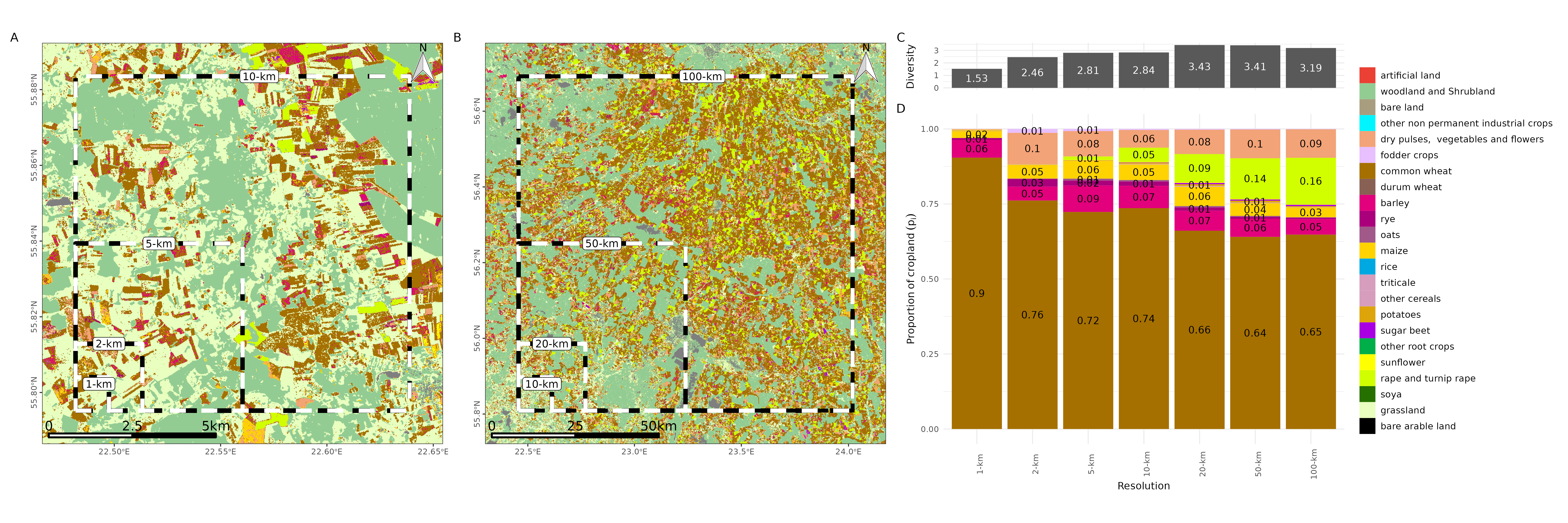}
        \caption{Latvia-Lithuania (N3700000E5100000 )}
        \label{fig:subfig3}
    \end{subfigure}
    
    \caption{Crop diversity is computed at different grid scales ranging from 1 to 100 km as illustrated here for a region in Latvia-Lithuania. For each subfigure, panel A shows the scales 1-km, 2-km, 5-km and 10-km with the crop map in background. Panel B shows the scales 10-km, 20-km, 50-km and 100-km with the crop map in background. Panel C shows the γ-diversity for the respective sample squares shown in A and B. Panel D shows the proportion of crop types for the different scales. These proportions are used to compute the Shannon diversity. }
    \label{fig:Fig-B-examnded}
\end{figure*}

\begin{figure*}[]
    \centering 
    \includegraphics[width=1\textwidth]{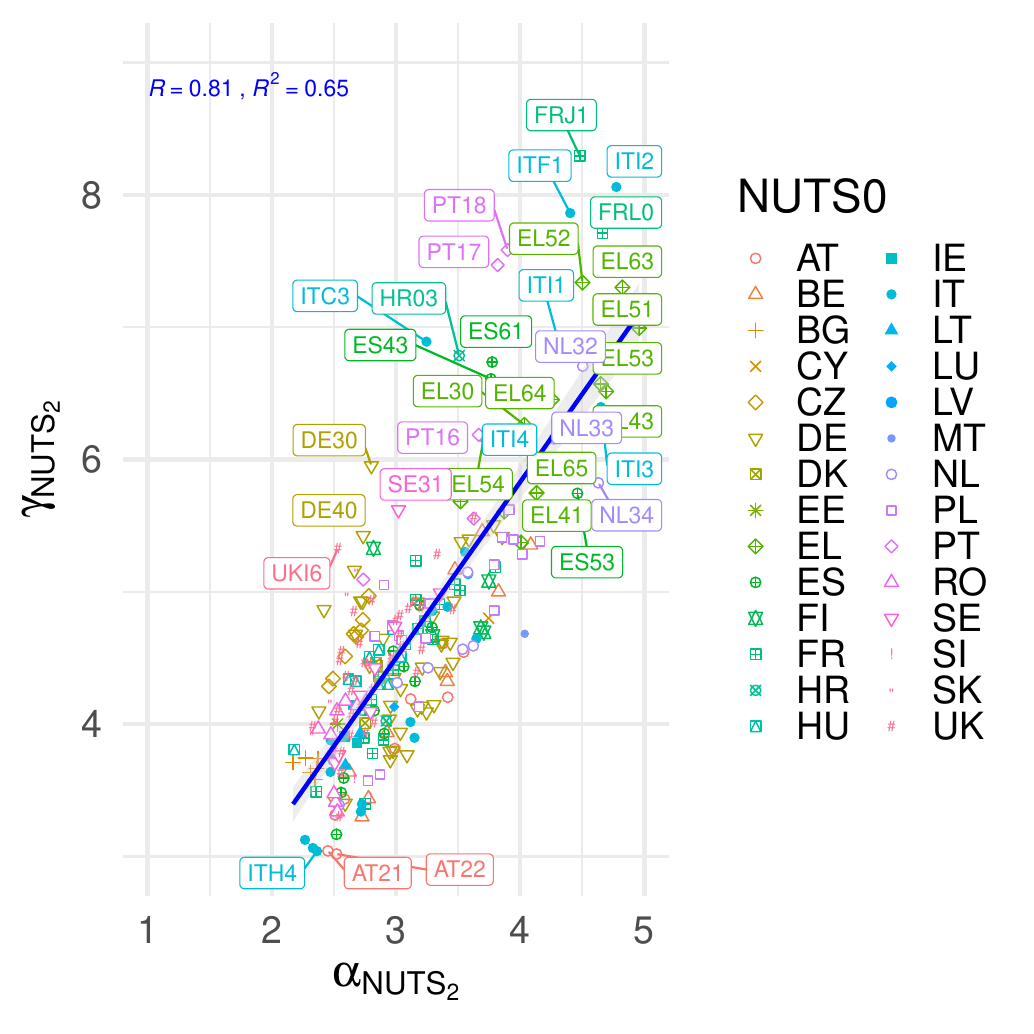} \vspace*{-.5cm}
    \caption{Fit of α-diversity ($\alpha_{NUTS_{2}}$) against γ-diversity ($\gamma_{NUTS_{2}}$) at sub-national (NUTS 2) level (see country abbreviations in \ref{tab:countrycodes})}
    \label{fig:Fig-Kb-bis}
\end{figure*} 

 \begin{figure*}[h]
    \centering 
    \includegraphics[width=1.\textwidth]{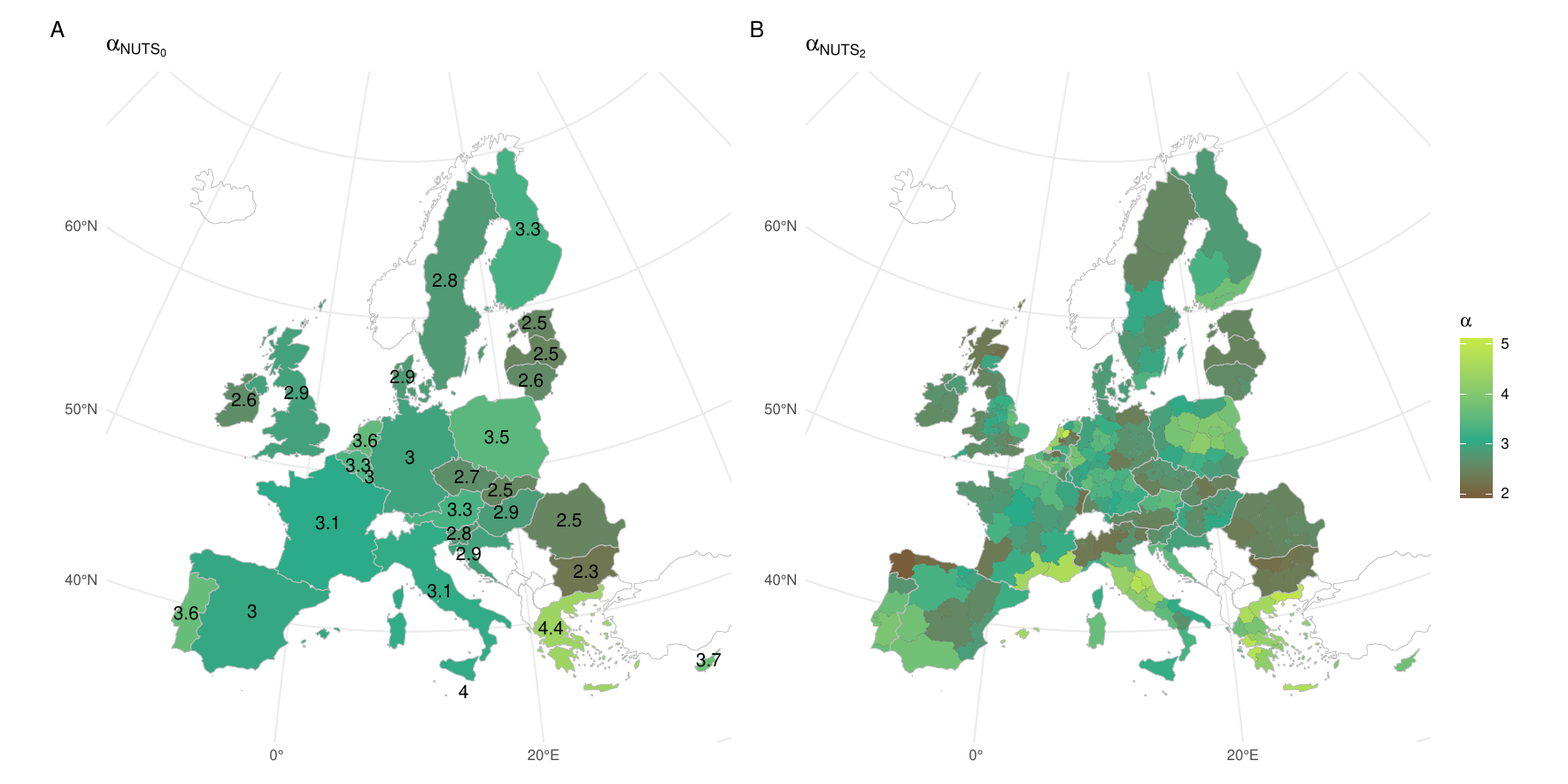} \vspace*{-.5cm}
    \caption{α-diversity at national (NUTS 0) and sub-national (NUTS 2) scale.}
    \label{fig:Fig-E}
\end{figure*}

\begin{figure*}[h]
    \centering 
    \includegraphics[width=1\textwidth]{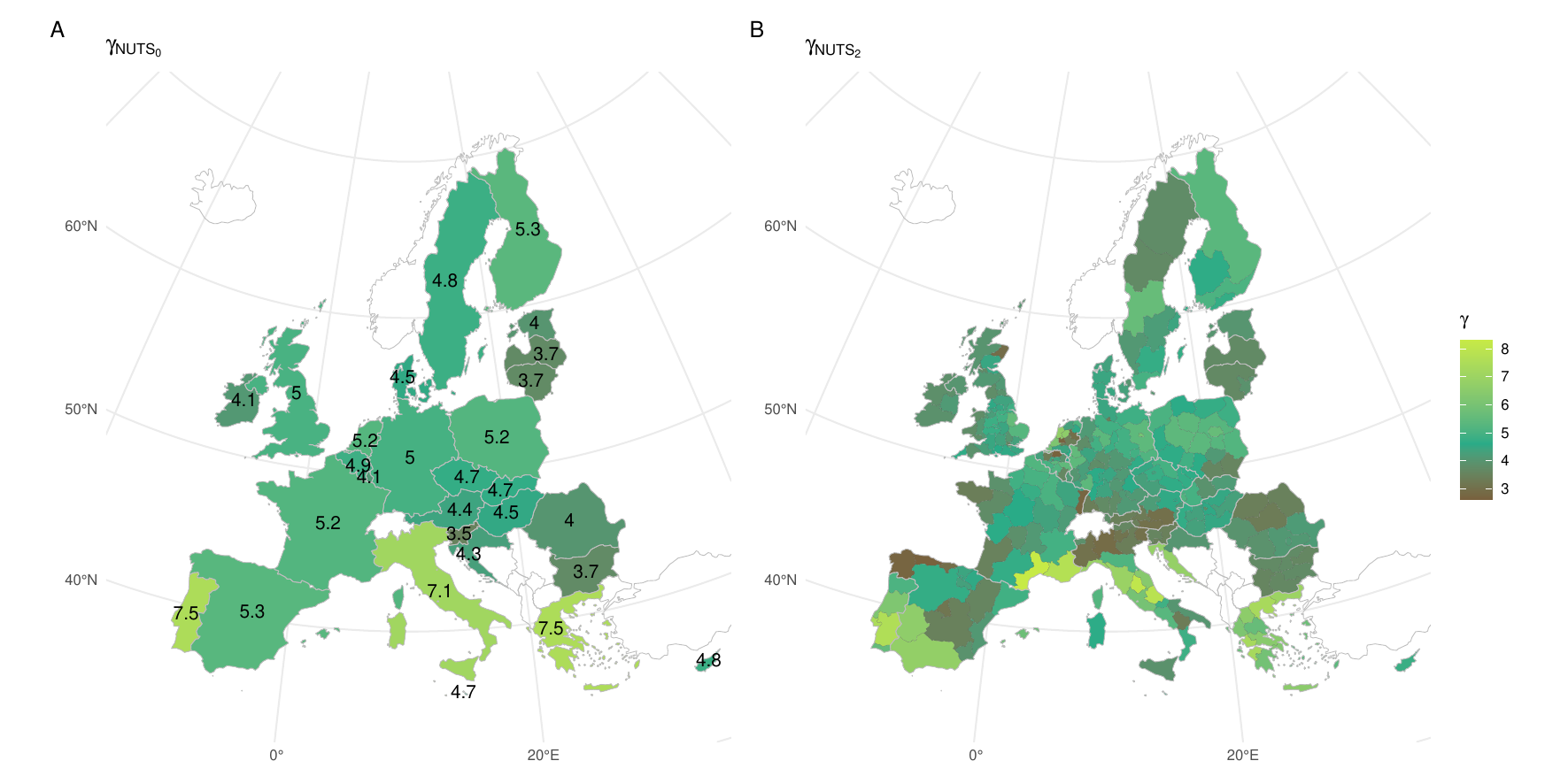} \vspace*{-.5cm}
    \caption{ γ-diversity at national (NUTS 0) and sub-national (NUTS 2) scale.}
    \label{fig:Fig-H}
\end{figure*}

\begin{figure*}[h]
    \centering 
    \includegraphics[width=1\textwidth]{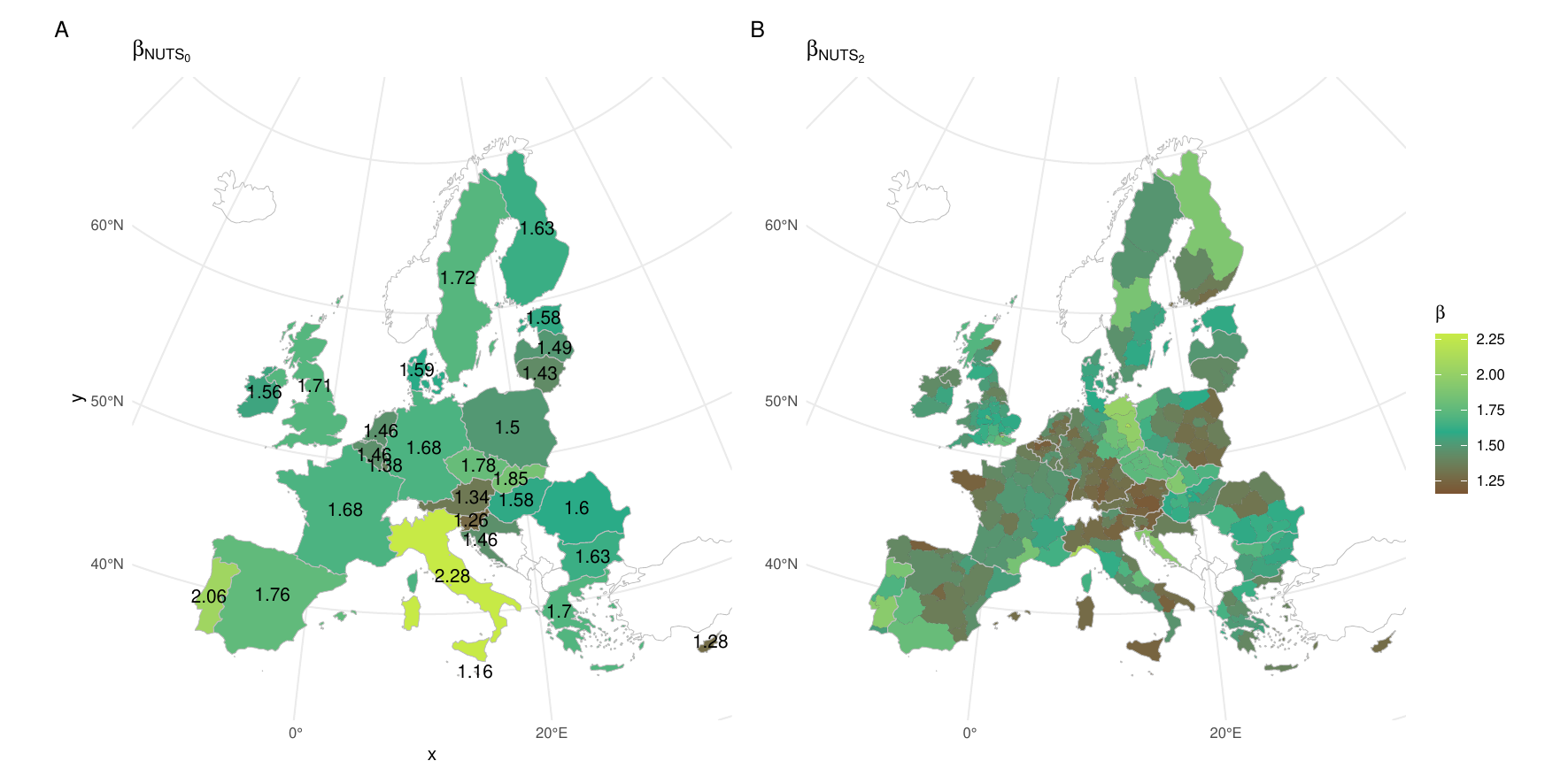} \vspace*{-.5cm}
    \caption{β-diversity at national (NUTS 0) and sub-national (NUTS 2) scale.}
    \label{fig:Fig-Kb}
\end{figure*} 

\begin{figure*}[!t]
    \centering 
    \includegraphics[width=0.9\textwidth]{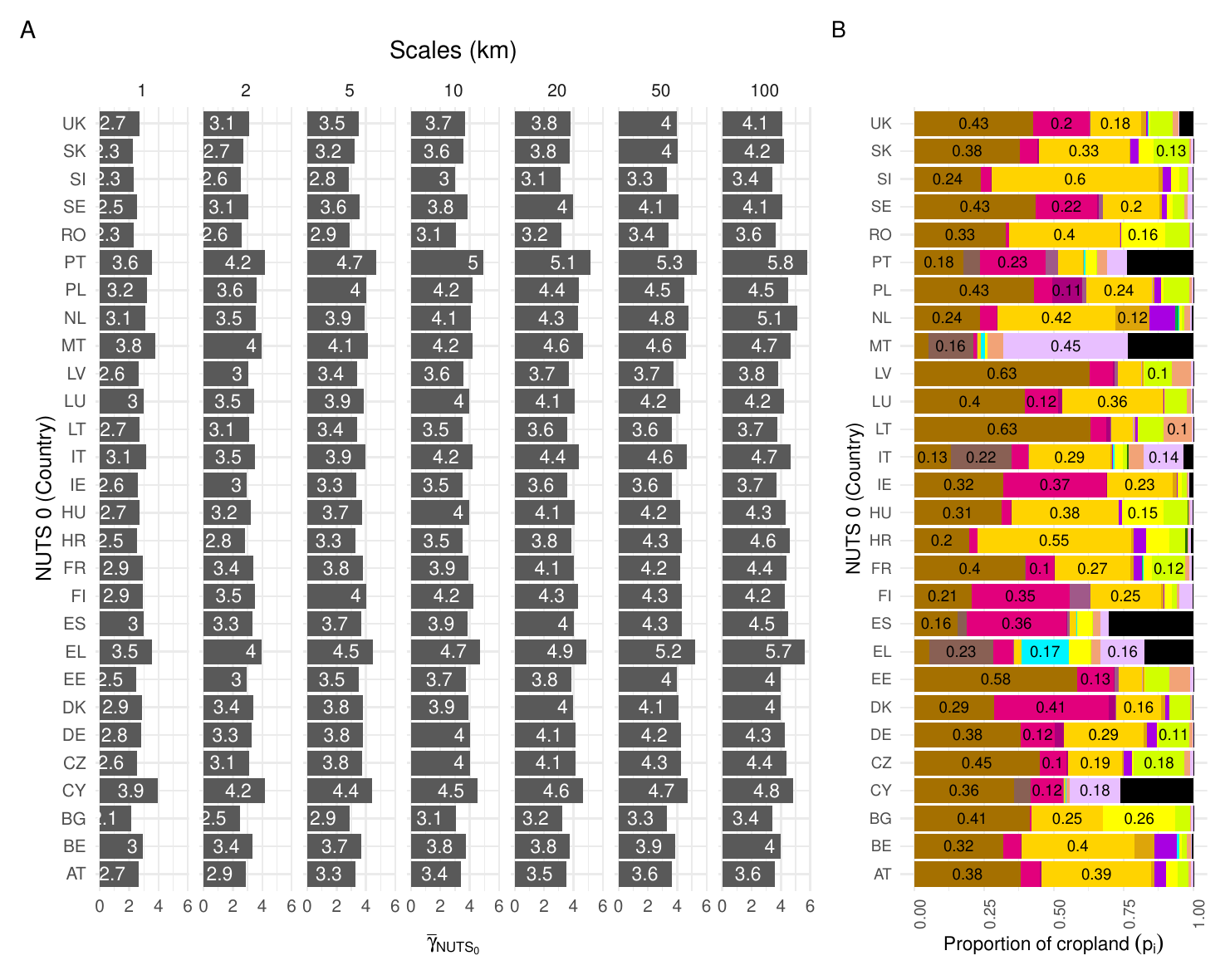} \vspace*{-.5cm}
    \caption{(A) γ-diversity by grid scale (1 km to 100 km) averaged at country level (NUTS0) and (B) corresponding crop type proportion ($p_i$) by country (NUTS 0). The crop type is only represented when  $p_i > 10\%$. Legend of crop type color mapping is shown on \ref{fig:Fig-B}}
    \label{fig:Fig-D-GammaGridValues}
\end{figure*} 
\begin{figure*}[!t]
    \centering 
    \includegraphics[width=.85\textwidth]{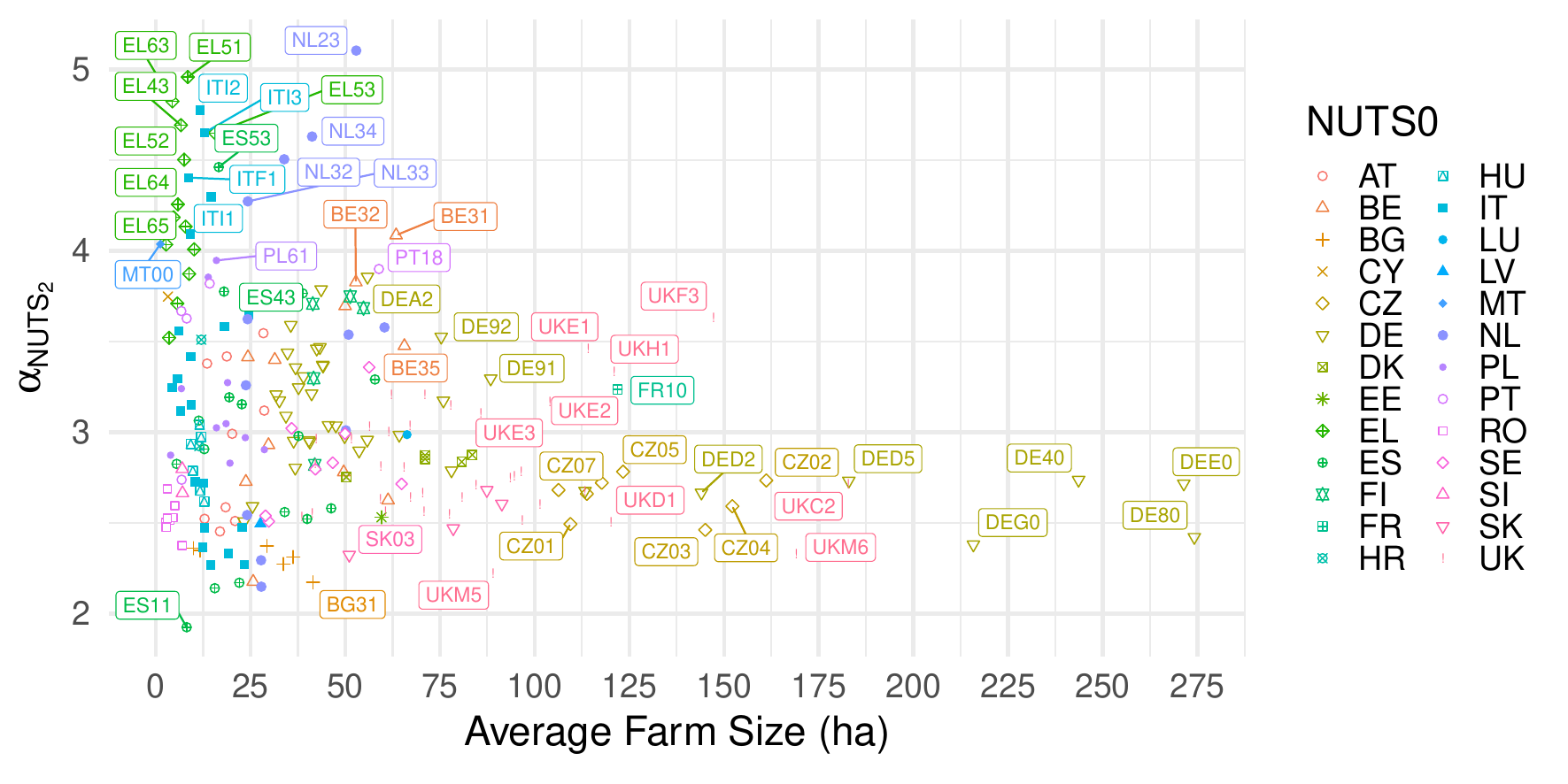}
    \caption{α-diversity as a function of average farm size (in hectares) at sub-national level (NUTS 2) in EU-28.}
    \label{fig:Fig-N-fss-diversity-n2}
\end{figure*}

\end{document}